\pdfoutput=1
\documentclass[iop]{emulateapj}
\usepackage{graphicx, subfigure}
\usepackage{amsmath}
\usepackage{listings}

\usepackage{xspace}
\usepackage{epstopdf}
\usepackage{url}

\usepackage{times}
\usepackage[english]{babel}
\usepackage{mathrsfs}

\usepackage{color}
\usepackage{xcolor}
\definecolor{green}{HTML}{33CC33}
\definecolor{red}{HTML}{FF3300}
\definecolor{blue}{HTML}{3333FF}
\usepackage[colorlinks=true,citecolor=blue,urlcolor=green]{hyperref}
\usepackage{mathrsfs}

\usepackage{relsize}

\usepackage{natbib,twoopt}
\usepackage[varg]{txfonts}

\usepackage[normalem]{ulem}

\renewcommand{\eqref}[1]{Equation~\ref{#1}}
\newcommand{\fref}[1]{Figure~\ref{#1}}
\newcommand{\sref}[1]{Section~\ref{#1}}

\newcommand{\ie}{i.\,e.} 
\newcommand{\eg}{e.\,g.} 

\newcommand{\pipe}{\ensuremath{\rm K2P^2}\xspace}
\newcommand{\numax}{\ensuremath{\nu_{\rm max}}\xspace}
\newcommand{\kp}{\emph{Kepler}\xspace}
\newcommand{\Kp}{\ensuremath{\rm Kp}\xspace}
\newcommand{\Kpt}{\ensuremath{\rm\tilde{ K}p_{1}}\xspace}
\newcommand{\Kptt}{\ensuremath{\rm\tilde{ K}p_{2}}\xspace}

\numberwithin{equation}{section}

\makeatletter
\def\maketag@@@#1{\hbox{\m@th\normalfont\normalsize#1}}
\makeatother

\bibpunct{(}{)}{;}{a}{}{,} 

\shorttitle{K2 pixel photometry}
\shortauthors{Lund et al.}

\begin{document}

\title{\large K2P$^2$ --- a photometry pipeline for the K2 mission\vspace*{0.3cm}} 

\author{Mikkel~N.~Lund$^{1,2\star}$}
\author{Rasmus~Handberg$^{1,2}$}
\author{\\Guy~R.~Davies$^{2,1}$}
\author{William~J.~Chaplin$^{2,1}$}
\author{Caitlin~D.~Jones$^{2,1}$\\ \vspace*{0.4cm}}

\affil{$^1$Stellar Astrophysics Centre (SAC), Department of Physics and Astronomy, Aarhus University,\\ Ny Munkegade 120, DK-8000 Aarhus C, Denmark; $^{\star}$\href{mailto:mikkelnl@phys.au.dk}{mikkelnl@phys.au.dk}}
\affil{$^2$School of Physics and Astronomy, University of Birmingham, Edgbaston, Birmingham, B15 2TT, UK}


\begin{abstract}
With the loss of a second reaction wheel, resulting in the inability to point continuously and stably at the same field of view, the NASA \kp satellite recently entered a new mode of observation known as the K2 mission. The data from this redesigned mission present a specific challenge; the targets systematically drift in position on a ${\sim}6$ hour time scale, inducing a significant instrumental signal in the photometric time series --- this greatly impacts the ability to detect planetary signals and perform asteroseismic analysis. Here we detail our version of a reduction pipeline for K2 target pixel data, which automatically: defines masks for all targets in a given frame; extracts the target's flux- and position time series; corrects the time series based on the apparent movement on the CCD (either in 1D or 2D) combined with the correction of instrumental and/or planetary signals via the KASOC filter \citep[][]{2014MNRAS.445.2698H}, thus rendering the time series ready for asteroseismic analysis; computes power spectra for all targets, and identifies potential contaminations between targets.
From a test of our pipeline on a sample of targets from the K2 campaign 0, the recovery of data for multiple targets increases the amount of potential light curves by a factor of ${\geq}10$.

Our pipeline could be applied to the upcoming TESS \citep[][]{2014SPIE.9143E..20R} and PLATO 2.0 \citep[][]{2013arXiv1310.0696R} missions.

\end{abstract}



\section{Introduction}
\label{sec:intro}

K2 \citep[][]{2014PASP..126..398H} is the continuation of the nominal NASA \kp mission \citep[][]{2010Sci...327..977B,2010PASP..122..131G} which ended with the loss of a second reaction wheel in May 2013. The stability solution for the \kp satellite is to balance in an unstable equilibrium against the Solar photon pressure and correct rolls with thruster firings, while pitch and yaw is controlled by the two remaining reaction wheels; this strategy allows for observations in fields along the ecliptic plane, with an observing length per field of close to 80 days. This time span is known as a ``Campaign" (C), and is the analogue to the 3 month ``Quarters" (Q) used in the nominal \kp{} mission. In the nominal mission targets were designated using a \kp Input Catalogue (KIC) number, which has now been replaced by the Ecliptic Plane Input Catalogue (EPIC) number.

The systematic pointing drift in the K2 observations, from the adopted stabilisation of the spacecraft, calls for new light curve correction methods. One such has recently been proposed by \citet[][]{2014PASP..126..948V}, and use the positions on the CCD as a function of time to decorrelate the induced variations in the light curve. The larger fields around targets in K2 --- needed to account for the apparent movement of the target on the CCD --- and the increased crowding from pointing toward the ecliptic means that often many stars are found in a given frame. This, combined with the potential lack of aperture masks from the \kp team, necessitates the development of new methods to extract the flux and position of targets from custom apertures, and this in an efficient and robust manner.

The paper is structured as follows: In \sref{sec:LC} we describe the steps taken in our light curve construction, starting from raw Target Pixel Files (TPF) and going to the definition of pixel masks and extraction of target positions and light curves. \sref{sec:corr} pertains to the correction of the light curves from the time dependent movement on the CCD; here we describe both our version of the 1D self flat-fielding introduced by \citet[][]{2014PASP..126..948V} in \sref{sec:corr1}, and our suggestion for a 2D approach in \sref{sec:corr2}.
In \sref{sec:c0ana} we present results from a test of our pipeline on a target sample during C0, and conclude in \sref{sec:dis}.


\section{Light curve construction}
\label{sec:LC}

The nominal \kp{} mission delivered a pixel aperture (a mask) where the chosen pixels optimised the mean signal-to-noise ratio (S/N) based on estimates of the pixel response function (PRF) and information from the KIC \citep[][]{2010ApJ...713L..97B,2010ApJ...713L..87J}.
This mask could be used to construct custom masks by adding or removing pixels to the starting mask based, for example, on the amount of flux in the pixels. This procedure was adopted in the KASOC filter pipeline \citep[][]{2014MNRAS.445.2698H} using the routine developed by Mathur et al. (in preparation). Masks are no longer delivered, at least not for the data releases made to date, which calls for a new method to define pixel masks. Masks constructed from ranking pixels in order of their S/N, and then including the number of pixels which optimises, for instance, the combined differential photometric precision (CDPP) noise metric \citep[][]{2011ApJS..197....6G,2012PASP..124.1279C} or the mean S/N could run into problems if signals from other stars are not removed; this is especially difficult if there are secondary objects in close proximity to the primary target. 

In the following we describe our pipeline for the construction of light curves, called \pipe (K2-Pixel-Photometry), which delivers both the position and flux for all the objects in the delivered frames. The stellar position as a function of time is used to filter the light curve from variations in flux induced by the movement of the stars over different pixels which have varying sensitivities (see \sref{sec:corr}). 
We define fixed masks from a summed image (see \sref{sec:sumimg}), which is large enough to encompass the stellar movement on the CCD. We go through the different steps in the \pipe pipeline below. In all examples times will be in truncated Barycentric Julian Date (TBJD)\footnote{this differs from the Barycentric \kp Julian Data (BKJD) given as $\rm BJD - 2454833$.} given as $\rm BJD - 2400000$.


\subsection{Background estimation}
\label{sec:bg}

As the initial step of \pipe{} we estimate the sky background as a function of time, because this contribution is unaccounted for in the flux from the raw K2 target pixel data. For each time step we calculate the mode of the flux kernel density estimation (using Scott's \citep[][]{SCOTT01121979} rule for setting the bin width) from all pixels as the maximum likelihood estimator for the sky background. We thus assume a uniform background flux across a given image.
\begin{figure}
\centering
\includegraphics[scale=0.44]{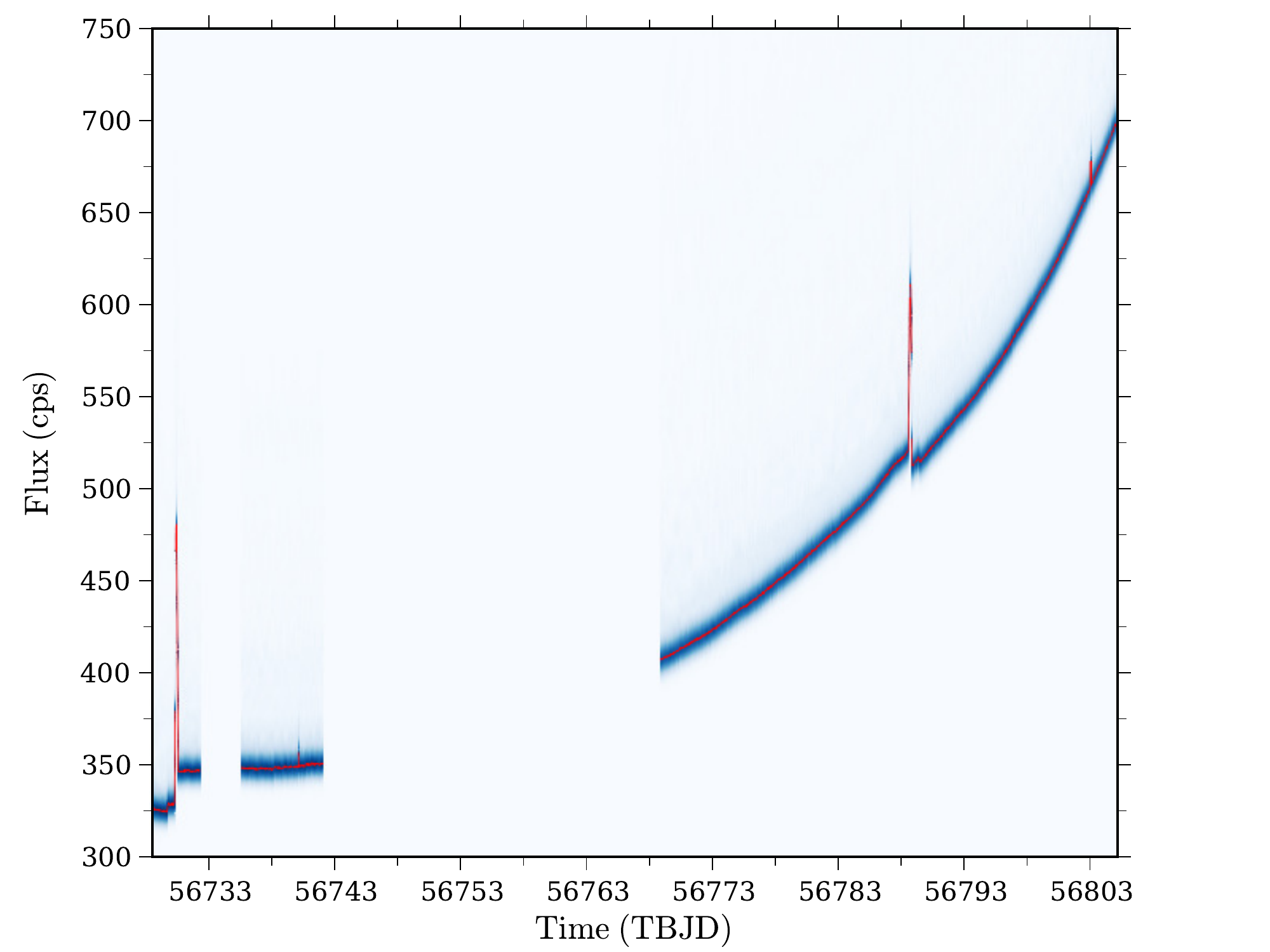}
\caption{Kernel distribution of flux within the pixel frame as a function of time during C0 (here for EPIC 202062417). The color scale goes from light for a low flux level to dark for a high flux level; the red line indicates the distribution mode. Times with quality flags indicating contamination of any sort have been excluded \citep[][]{kepman}. The presence if a ghost image of Jupiter elevates the background flux between $56728-56788$ TBJD, with high flux-spikes at the beginning and end of this interval where Jupiter enters and exist the focal plane and making specular reflections.}\vspace{1em}
\label{fig:background}
\end{figure}

The sky background level is far from constant, but increases gradually (by around ${\sim}25\%$ in C0) over the course of a campaign; a typical example can be seen in \fref{fig:background}. In C0 the background level was further increased for many channels by the antipodal ghost image of Jupiter as it fell on one of \emph{Kepler's} dead modules\footnote{\url{http://keplerscience.arc.nasa.gov/K2/C0drn.shtml}}. The change in background levels can largely be attributed to changing levels of stray light entering the photometer from the change in angle between the Sun and the photometer, and is thus additive. Secondary changes might come from changes in focus as the heating of the spacecraft varies.

If the background level variation is unaccounted for it will appear in the extracted light curve; it is preferential to isolate this component and separate it from trends caused by the degraded attitude.


\subsection{Summed image}
\label{sec:sumimg}

\begin{figure}
\centering
\includegraphics[scale=0.44]{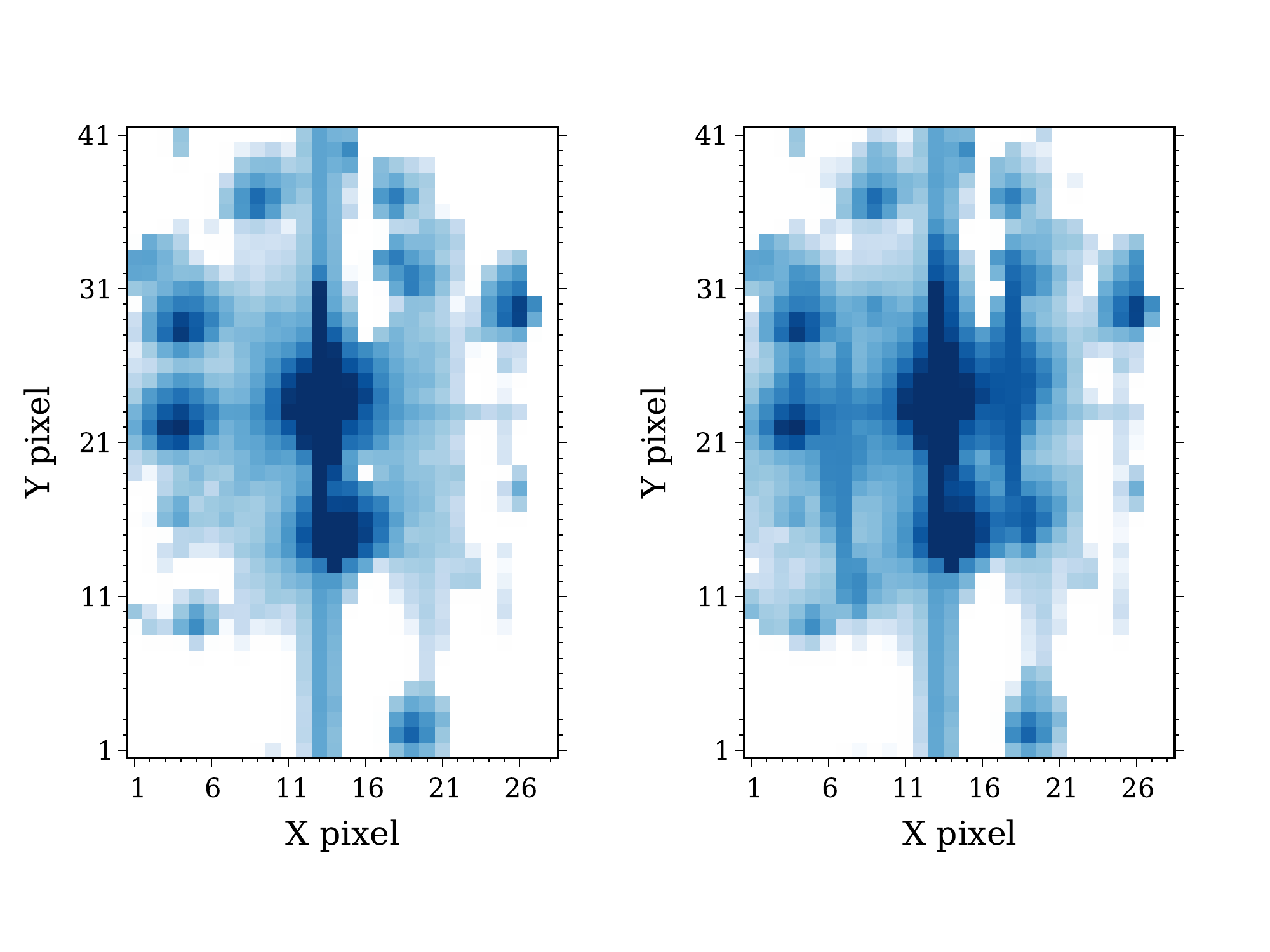} \vspace{-3em}
\caption{Summed image from short-cadence C0 data of EPIC 202062417, with the background mode (see \fref{fig:background}) subtracted from individual frames. Colour scale is on a logarithmic scale going from light (low flux) to dark blue (high flux), and negative levels are truncated to 0. Left: Summed image with frames having a bad quality flag removed. Right: Summed image with all frames included; here are ghost image of the brightest targets appear shifted approximately two pixels up and five pixels to the right.	}
\label{fig:sumimg}
\end{figure}

For setting pixel masks we create a summed image. Here, frames are co-added after first having subtracted the corresponding sky background levels (see \sref{sec:bg}). We make use of the quality flags available in the pixel data fits files \citep[][]{kepman}, and ignore all frames with a flag indicating any non-optimal data. The effect of neglecting this is illustrated in \fref{fig:sumimg}. Including frames with bad quality flags, for instance when reaction wheel momentum dumps are made, results in the creation of a shifted ghost image. If a summed image including a shifted ghost image would be used in setting masks, these would be much larger than needed and would essentially only add noise for the majority of the time series. It would also be difficult for an automatic routine, that can separate close targets, to identify the ghost image as belonging to the main target rather than being a target in its own. 


\subsection{Pixel mask selection}
\label{sec:pixelmask}
To fix the masks we first select which pixels can be included in a mask by setting a flux threshold.
The threshold is obtained as the median absolute deviation (MAD) of the summed image flux distribution which falls to the left-hand-side of the mode of the distribution. Only the left-hand side of the distribution is used as the right-hand side is influenced more strongly by the stellar flux.
 
On the pixels with flux levels above the threshold we run an unsupervised clustering algorithm to locate targets in the frame and set individual masks for these.
Specifically, we use the \emph{density-based spatial clustering of applications with noise} routine \citep[DBSCAN;][]{ref:dbscan} as implemented in the \texttt{Python}-based library Scikit-learn\footnote{\url{http://scikit-learn.org/}} \citep[see][]{paper:scikit-learn}.
DBSCAN only takes two input parameters: a neighbourhood radius $r_c$, and a minimum number of points needed to form a cluster $N_{\rm min}$. 
Given the regularity of the pixel grid, these parameters can be set optimally \textit{a priori} to yield a desired output. An advantage of the DBSCAN routine is that it does not need a predefined number of clusters, and that the clusters can have very irregular shapes --- allowing it to encompass the spatial distribution of flux from a star on the CCD in K2, which depends both on time and position on the focal plane.

The working principle of the DBSCAN is, briefly: (1) select at random a point, with ``points'' being the pixels with flux above the threshold; (2) check how many other points $N_c$ are within the neighbourhood radius $r_c$ of the selected point; (3) if $N_c \geq N_{\rm min}$ the point is designated as a \emph{core point} and the start of a cluster, otherwise, if $N_c < N_{\rm min}$, it is (at this step) designated as a \emph{noise point}; (4) step (2) is now run on points within $r_c$ of the first point, and so on for their respective neighbourhood points, and points are added to the first cluster until no more points are \emph{density reachable} --- that is, can be connected by a chain of points to the initial point seeding the cluster; (5) a point that falls within $r_c$ of a cluster core point, but which has $N_c < N_{\rm min}$ in its own neighbourhood, is designated as an \emph{edge point} to the cluster. Note that if such an edge point was the first considered by the routine, it would have been flagged as a noise point, but it will change status later in the routine if found within $r_c$ of a cluster core point; (6) when no more points can be added to the first cluster, one of the remaining points is selected at random and the steps are run through anew. This continues until all points have a designation.

\begin{figure}
\centering
\includegraphics[scale=0.44]{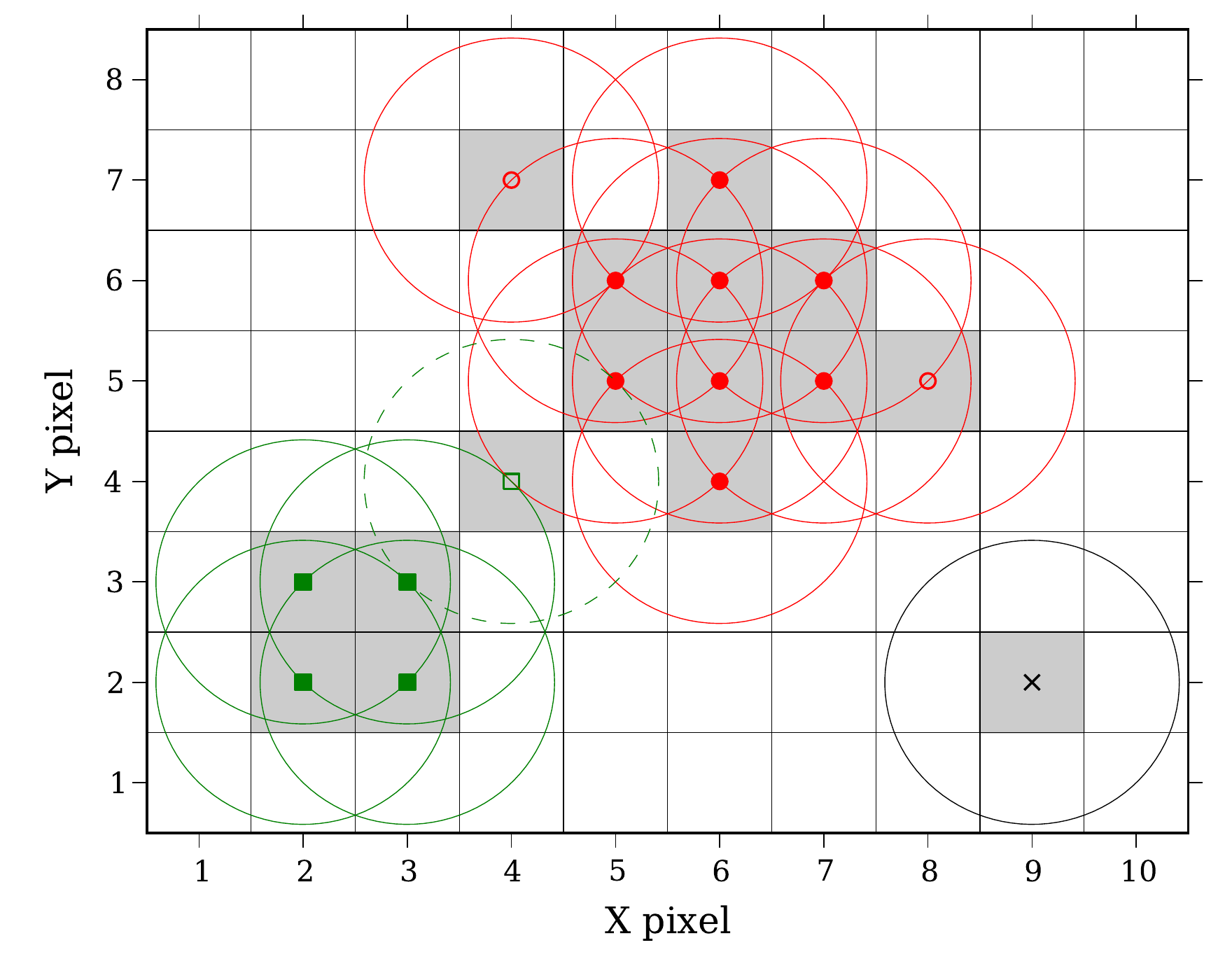}
\caption{Illustration of the working principle of the DBSCAN clustering algorithm. Pixels with flux above a pre-set threshold that are used in the clustering are marked in gray. Two clusters are identified with constituent pixels marked by either red circles or green squares; the cross marker gives the pixel identified as noise, and for each pixel its neighbourhood radius, $r_c$, is indicated. Filled markers indicate \emph{core} cluster members; empty markers give the \emph{edge} members. The edge member that could belong to either cluster (in this case associated with the green cluster) is indicated with a dashed neighbourhood radius. }
\label{fig:dbscan}
\end{figure}

An illustration is provided in \fref{fig:dbscan} where we set $r_c=\sqrt{2}$ pixels and $N_c=3$.
Each of the clusters returned are seen as a target, with the core and edge members of the individual clusters defining the outer boundary of the masks of the targets.
Edge members within reach of more than one cluster could belong to either one of the clusters, and the membership of such a point would be determined entirely by the random initialisation of the routine. Core members, on the other hand, can be assigned clusters with full determinism, and will always group in the same way. We find, however, that the gain from a larger mask, which includes both core and edge members, outweighs the potential ambiguity and loss of repeatability from including a point that could belong to more than one cluster. In order to make the clustering reproducible we chose a fixed random seed for the algorithm\footnote{specifically we used the seed 1138 \citep[see][]{GLucas1977}.}, which ensures that the clustering and designation of point will stay the same for a rerun with the same settings.
\begin{figure*}
\centering
\includegraphics[scale=0.45]{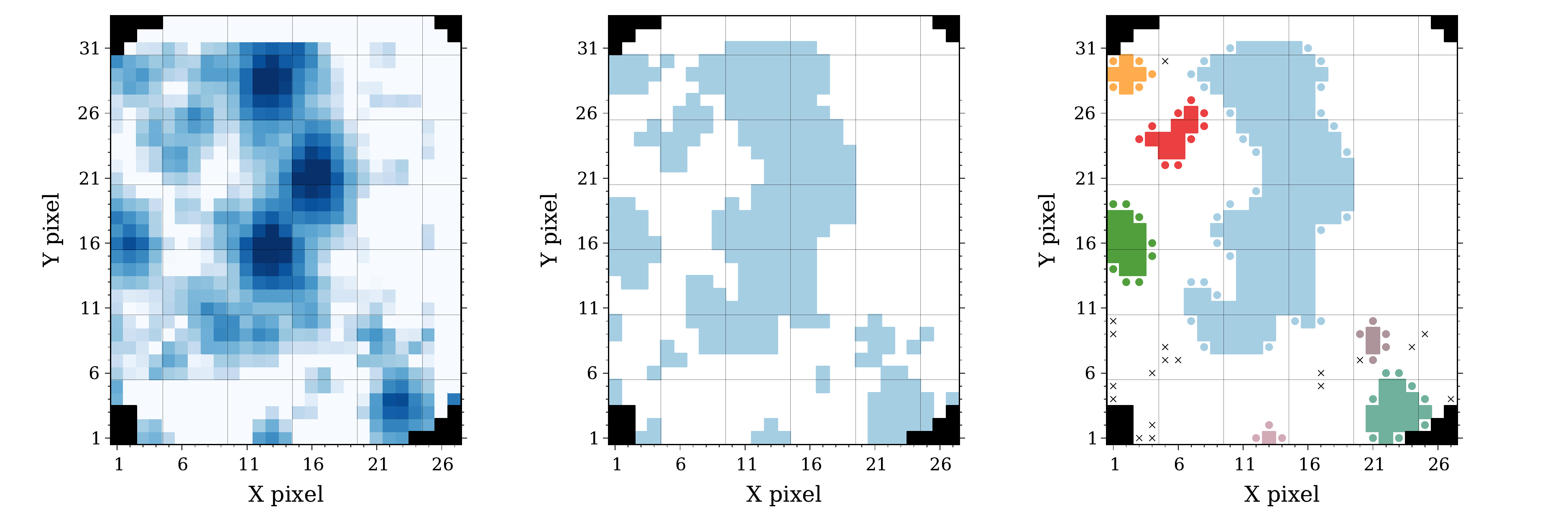}
\caption{Illustration of initial steps (Section~\ref{sec:sumimg}-\ref{sec:satura}) in selection of pixel masks (here for C0 observations of EPIC 202127012). For pixels marked in black no flux is collected. Left: Summed image (\sref{sec:sumimg}) with the background mode (\fref{fig:background}) subtracted from individual frames; the colour scale is on a logarithmic scale going from light (low flux) to dark blue (high flux), and negative levels are truncated to 0. Middle: Collection of pixels with flux levels above a predefined threshold (\sref{sec:pixelmask}). Right: Clusters identified from running the DBSCAN clustering algorithm, each of which is marked with a distinct colour; filled pixels mark core members; circles indicate edge members, and crosses give pixels identified as noise. In this run we used $r_c=1$ pixels and $N_c=3$.}
\label{fig:pro1}
\end{figure*}
\begin{figure*}
\centering
\includegraphics[scale=0.45]{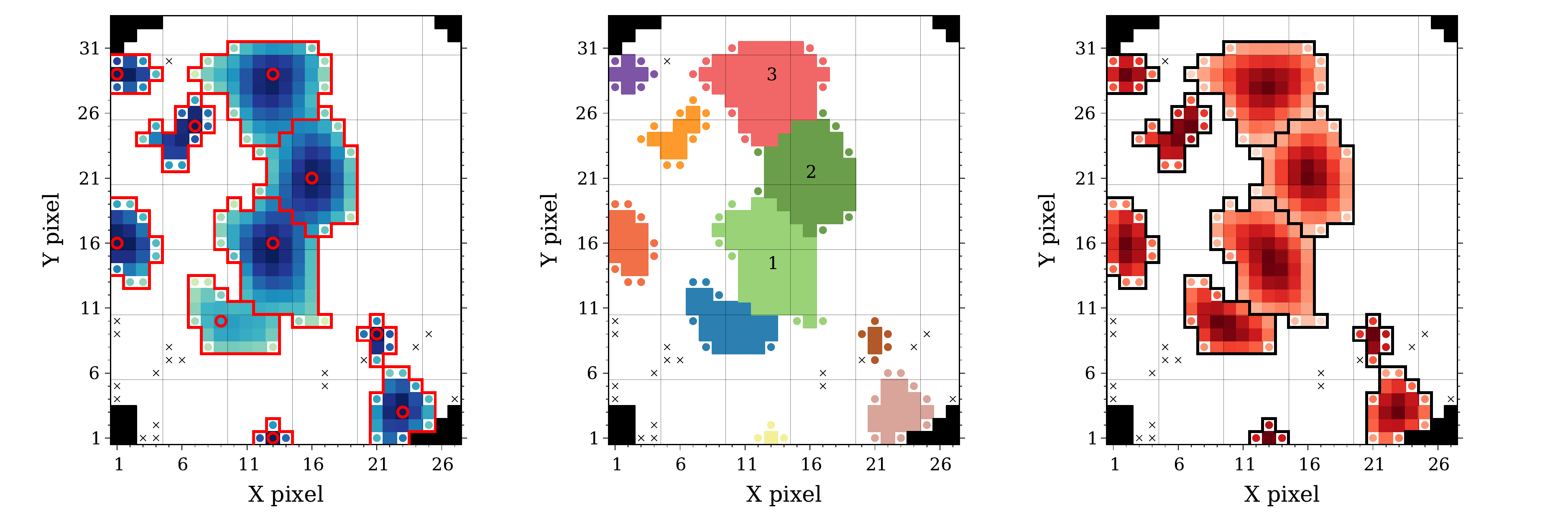}
\caption{Illustration of final steps (Section~\ref{sec:waters}-\ref{sec:pos_flux}) in setting pixel masks and extraction of target positions and fluxes (as in \fref{fig:pro1} for C0 observations of EPIC 202127012). For pixels marked in black no flux is collected. Filled pixels mark core members, while circles indicate edge members, and crosses give pixels identified as noise. Left: Application of the watershed segmentation algorithm on the clusters identified in the right panel of \fref{fig:pro1}; the colour scale indicate the relative negative flux level for each cluster individually (\ie, level do not translate between clusters) after application of a Gaussian 2D filter. Levels go from light (low negative flux) to dark blue (high negative flux) and are rendered on a logarithmic scale. Red circles show the identified local minima which are used as markers in the watershed routine. Red lines give the mask borders after the watershed segmentation, as seen the large central cluster has been divided into four components. Middle: Masks of the now ten identified targets, each rendered in a different colour. The three brightest targets have been designated with numbers; the primary target is star no. 1 (see \fref{fig:wcs}). Right: An example of weights ($w_i$) of pixels within the different masks, here given by the euclidean distance between a given pixel to the nearest pixel outside the mask; the scale is again only applicable for the individual masks and do not translate between masks. Black lines indicate the mask borders.}
\label{fig:pro2}
\end{figure*}


\subsection{Saturated targets}
\label{sec:satura}
The setting of masks for saturated targets calls for some extra attention. The saturation limit is at a \kp magnitude\footnote{nearly equivalent to an R band magnitude \citep[][]{2010ApJ...713L..79K}.} of $\Kp\sim 11.3$ \citep[][]{2010ApJ...713L.160G}, and saturated targets will typically have pixel column trails along which flux spills, or bleeds. If the ends of these trails fall outside the mask the variability in the flux will be missed, resulting in a high-flux truncation of the light curve. 
The bleed-out is position dependent from the varying pixel sensitivities across the focal plane, but in K2 it will also depend on time, because the targets now have a time and position dependent movement on the detector. This results in a even poorer predictability of the amount of bleed-out; we find that bleed-outs generally start for $\Kp \lesssim 9$.
\begin{figure}[h!]
\centering
\includegraphics[scale=0.4]{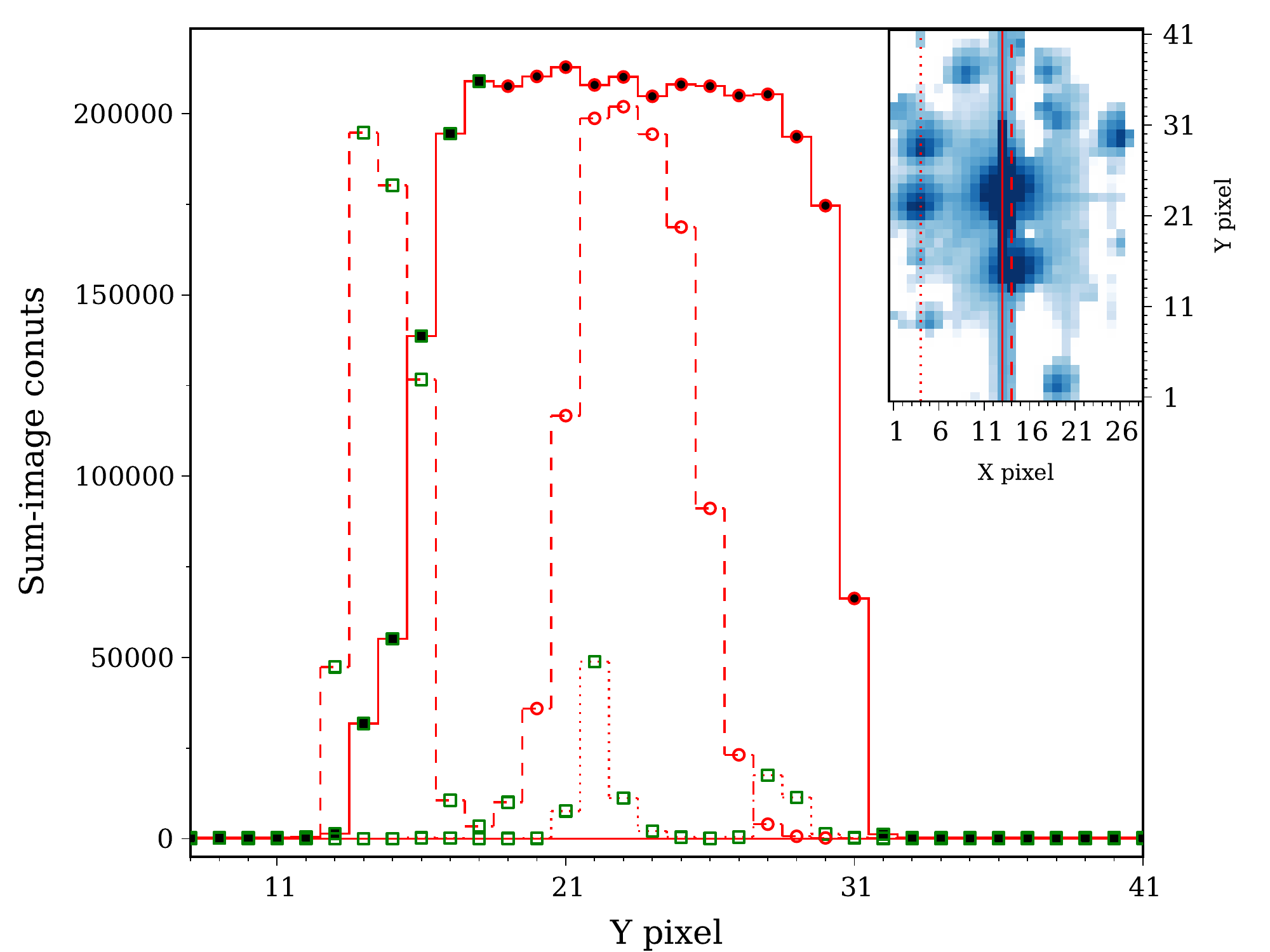}
\caption{Addition of pixels in bleed-out columns for the saturated target EPIC 202062417, using short-cadence C0 data. Different lines correspond to different X-pixel columns, see insert. Red circle (green square) markers indicate pixels in (not in) the mask of the main target (see \fref{fig:sumimg}). Markers filled with black gives the pixels included in the mask after check for saturation bleed-out; in the given case the whole of column 13 is added to the mask of the primary target.} 
\label{fig:saturated}
\end{figure}

An optimum inclusion of bleed-out trails is particularly difficult if the trail extends to other targets, or reaches the detector edge; in such cases a trade-off must be made between the amount of flux that can be included from the main target and the contamination from neighbouring targets.

We have implemented the following procedure for dealing with saturated targets (see \fref{fig:saturated}): For a given target we compute for each pixel-column, using pixels in the target's mask, the ratio between the absolute value of the median of the first differences in the flux counts of the pixels and the maximum flux count of the pixels. A low value of this ratio indicates a small relative variability in the flux counts, as would be the case for a near-constant flux level in a column with many saturated pixels. If the ratio is below $1\%$, and the median of the pixel flux counts (still only for the pixels in the mask) is equal to or larger than half of the maximum flux count for the entire mask, the column is taken as having saturated pixels. The restriction on the median of the flux counts ensures that columns containing many pixels with flux levels close to the background, where the relative variability also is small, are recognised as non-saturated. For the columns identified as saturated we then add pixels to the mask if these have counts above the flux threshold used in \sref{sec:pixelmask}. This could potentially result in pixels belonging to both a saturated target as well as a nearby secondary target.

For the brightest and most saturated targets ($\Kp \lesssim 8$), with bleed-outs spanning many tens of pixels (\eg, EPIC 202061312), with much flux contained in diffraction spikes on the CCD, and typically with multiple secondary targets in the near vicinity the mask should be defined manually --- as was done, for instance, for the 16 Cyg stars in the nominal \kp mission \citep[][]{2014ApJ...782....2L, 2015MNRAS.446.2959D}.


\subsection{Separating close targets}
\label{sec:waters}

After a set of clusters has been identified there is still the possibility that a given cluster might encompass two or more stars if these lie close to each other. To separate such targets in a given cluster we run an algorithm often used in image segmentation problems known as the \emph{watershed} method \citep[][]{citeulike:2335595, Beucher1993}, as implemented in Scikit-image\footnote{\url{http://scikit-image.org/}} \citep[see][]{ref:scikit-image}. The idea in a watershed algorithm is to find the line(s) between two or more regions, that may be seen as topographical surfaces; considering two neighbouring catchment basins that are flooded with water, the watershed will be the line where water levels meet.

To transform the pixel clusters to a topographical relief each point in a given cluster is assigned a value from the metric given either by the negative of the euclidean distance to the nearest background point (\ie, a point not in the specific cluster) or the negative value of its flux. This results in cluster points close to the edge having low negative values while central points of the cluster, which are further away from the background and generally have higher flux levels, have high negative values; this constitutes the catchment basins. If a cluster includes two or more stars that are not completely covered by a common envelope they will have distinctive central dips in both the distance and the flux metric. If the stars share a common envelope (seen if the stars are very close, or if one star greatly outshines the other) the flux metric is superior in making distinctive dips for the two (or more) stars; the distance metric will rather make a central dip for the whole region covered by the common envelope. As the default we use the flux metric to separate targets. 

In the adopted watershed algorithm we first identify the local minima of the metric used and then use these as markers for the centres of the catchment basins which are then flooded to find the watershed lines. To avoid noise peaks being considered as markers we first smooth the surface with a 2D Gaussian filter, and then locate the most prominent minima --- these are then fed as markers to the watershed routine. We now have pixel masks for all targets in a given frame.
\begin{figure}
\centering
\includegraphics[scale=0.4]{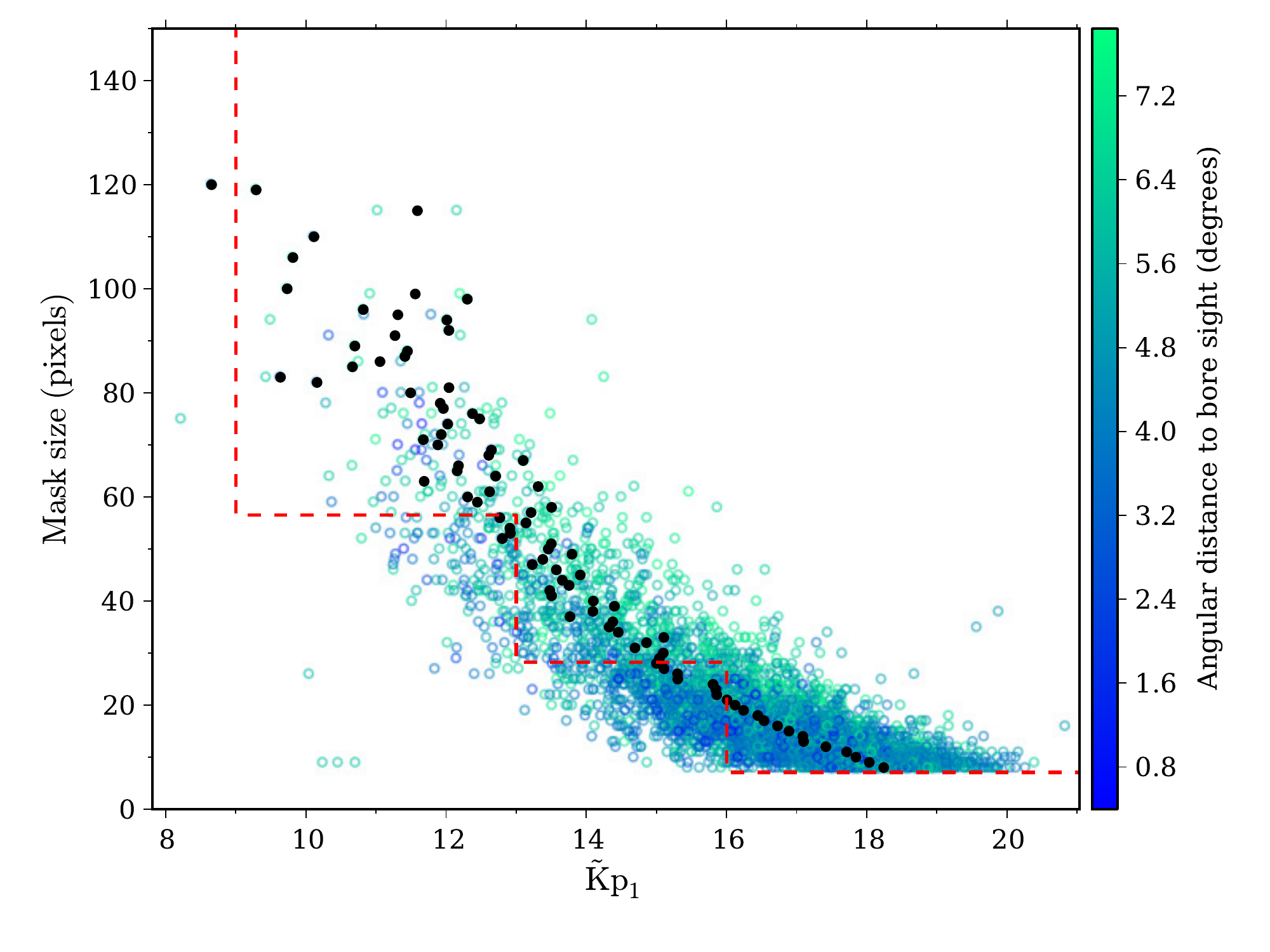}
\caption{Mask size as a function of the proxy \kp magnitude \Kpt (see \eqref{eq:mag}) for a sample of $4691$ C0 targets (see \sref{sec:c0ana}). Black markers indicate the median \Kpt for each of the discrete mask sizes; the colour coding indicate the angular distance for each target to the space craft bore sight; the red dashed line gives the mask sizes from \citet[][]{2014arXiv1412.6304A}.} 
\label{fig:size_vs_mag}
\end{figure}

Following the method outlined in Sections~\ref{sec:pixelmask}-\ref{sec:waters} for setting the pixel mask, we obtain for a sample of 4691 targets observed during C0 (see \sref{sec:c0ana}) mask sizes as a function of magnitude (see \sref{sec:magni}) as given in \fref{fig:size_vs_mag}. Here we note a slight gradient in the mask size as a function of angular distance to the space craft bore sight for a given \Kpt; this is as expected because the arc traced in apparent movement on the CCD from the roll of the spacecraft increases linearly with distance from the bore sight. The scatter in this relation will have contributions from the dependence of the degree of flux-smearing on the target position on the focal plane, and the uncertainty of the determined magnitude. For comparison we also show the magnitude dependence of aperture sizes from \citet[][]{2014arXiv1412.6304A}, where the authors use circular apertures/masks.


\subsection{Target magnitudes}
\label{sec:magni}
Our pipeline enables the extraction of data for multiple targets in a given frame, but from the information in the target pixel data we only have a \kp magnitude, \Kp, for the primary target. 
First, however, it should be noted that when targets were proposed for C0 the EPIC did not exist. Therefore, a magnitudes given in the EPIC\footnote{\url{http://archive.stsci.edu/k2/epic.pdf}} for a given C0 target is the one provided by the principal investigator proposing the target, rather than one computed by the \kp team. For the same reason no information is given in the \texttt{KepFlag} entry of the EPIC for C0, which is suppose to contain information on the data used to compute \Kp --- one should therefore consult the proposal of a given target to assess how the magnitude was constructed.
For the sample of targets we have analysed, viz., the proposal GO1038 (see \sref{sec:c0ana}), it turns out that the EPIC \kp magnitudes are given by $J$-band magnitudes from the Two Micron All Sky Survey \citep[2MASS;][]{2006AJ....131.1163S}. To transform these $J$-band magnitudes to more proper \kp magnitudes we use the transformation from \citet{2012ApJ...746..123H} between \Kp and 2MASS $J-K_s$ colors.

\begin{figure*}
\centering
\includegraphics[scale=0.4]{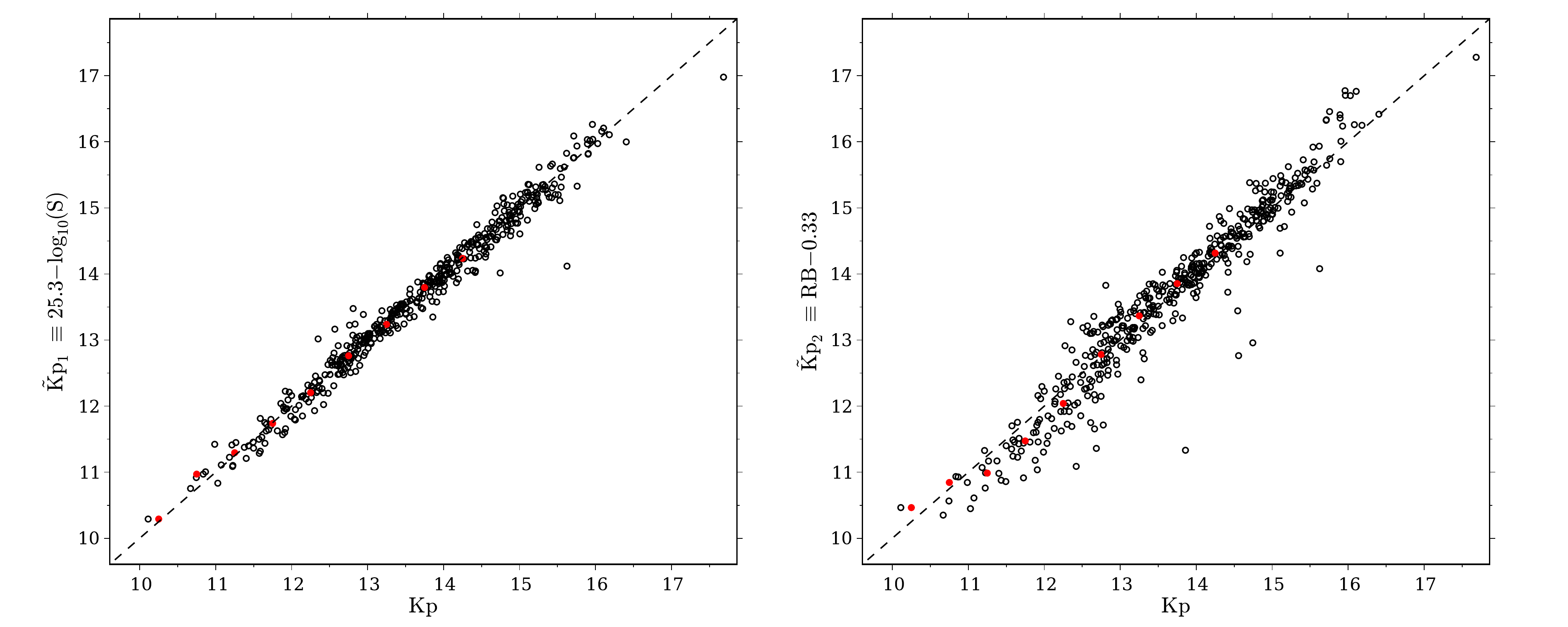}
\caption{Relation between proxy \kp magnitudes \Kpt (\eqref{eq:magrel1}) and \Kptt (\eqref{eq:magrel2}) and the nominal \kp magnitude, \Kp, computed from 2MASS $J-K_s$ colors. The dashed lines give the 1:1 relation. For 0.5 magnitude bins in \Kp the median \Kpt and \Kptt values are given by a red marker.} 
\label{fig:mag_vs_mag}
\end{figure*}

In order to investigate how parameters such as mask sizes and noise measures vary with magnitude we need a way to estimate \Kp for all targets in a given frame. We approximate \Kp by the proxy \kp magnitude \Kpt defined as:
\begin{equation}\label{eq:magrel1}
\Kpt \equiv 25.3 - 2.5\log_{10}(S)\, ,
\end{equation}
where ``$\rm S$'' denotes the median of the flux time series extracted for the target (in units of $\rm e^-/s$). The correspondence between \Kp and \Kpt is shown in the left panel of \fref{fig:mag_vs_mag}; some of the scatter in this relation will originate from the scatter in the mask size versus magnitude relation (see \fref{fig:size_vs_mag}), and a variation in pixel sensitivities between targets. We note that \citet[][]{2014arXiv1412.6304A} defines a proxy \kp magnitude in the same manner and also find an offset of ${\sim}25.3$. 
As a means of identifying targets falling within a given mask (see \sref{sec:loc}) we use the USNO-B1.0 catalogue \citep[][]{2003AJ....125..984M}, which is an all-sky catalogue with a completeness down to $\rm V=21$. We would like a measure of \Kp for all targets from the USNO-B1.0 catalogue within a given frame, because this is used in the identification of targets (see \sref{sec:loc}). In addition we can estimate potential contaminations when multiple targets fall within the same mask. 
For each of the identified targets from the USNO-B1.0 catalogue that fall within a given mask, we first compute the magnitude RB from the USNO-B1.0 R- and B-band magnitudes:
\begin{equation}\label{eq:mag}
\rm RB  =\left\{ 
\begin{array}{l l}
   \rm  0.1B_{mag} + 0.9R_{mag}\, ,  &\rm  \quad (B_{mag}-R_{mag})\leq 0.8\\
   \rm  0.2B_{mag} + 0.8R_{mag}\, ,  &\rm \quad (B_{mag}-R_{mag})>0.8\\
  \end{array} \right.
\end{equation}
According to \citet[][]{2011AJ....142..112B} this corresponds to the way \kp magnitudes, \Kp, are calculated in the KIC if only R- and B-band magnitude are available.
We define the following relation as a second proxy \kp magnitude:
\begin{equation}\label{eq:magrel2}
\Kptt \equiv RB - 0.33\, .
\end{equation}
The correspondence between \Kp and \Kptt is shown in the right panel of \fref{fig:mag_vs_mag}. 

The relation giving the \Kpt proxy has the smallest amount of scatter, and will be used to relate mask sizes and noise measures to magnitude; \Kptt will be used in the identification of targets, and estimation of contaminations.
An advantage of having both \Kpt and \Kptt is also that a large discrepancy between the two measures can be used to identify targets where the mask is either much too large or small. 

The offsets for both \Kpt and \Kptt were estimated in a Bayesian manner using the affine invariant \emph{emcee} sampler \citep[][]{2013PASP..125..306F}, and given by the median of the marginalized posteriors; the uncertainties were obtained from the 68$\%$ highest probability density of the marginalized posteriors.


\subsection{Locating main and secondary targets}
\label{sec:loc}

\begin{figure*}
\centering
\includegraphics[scale=0.40]{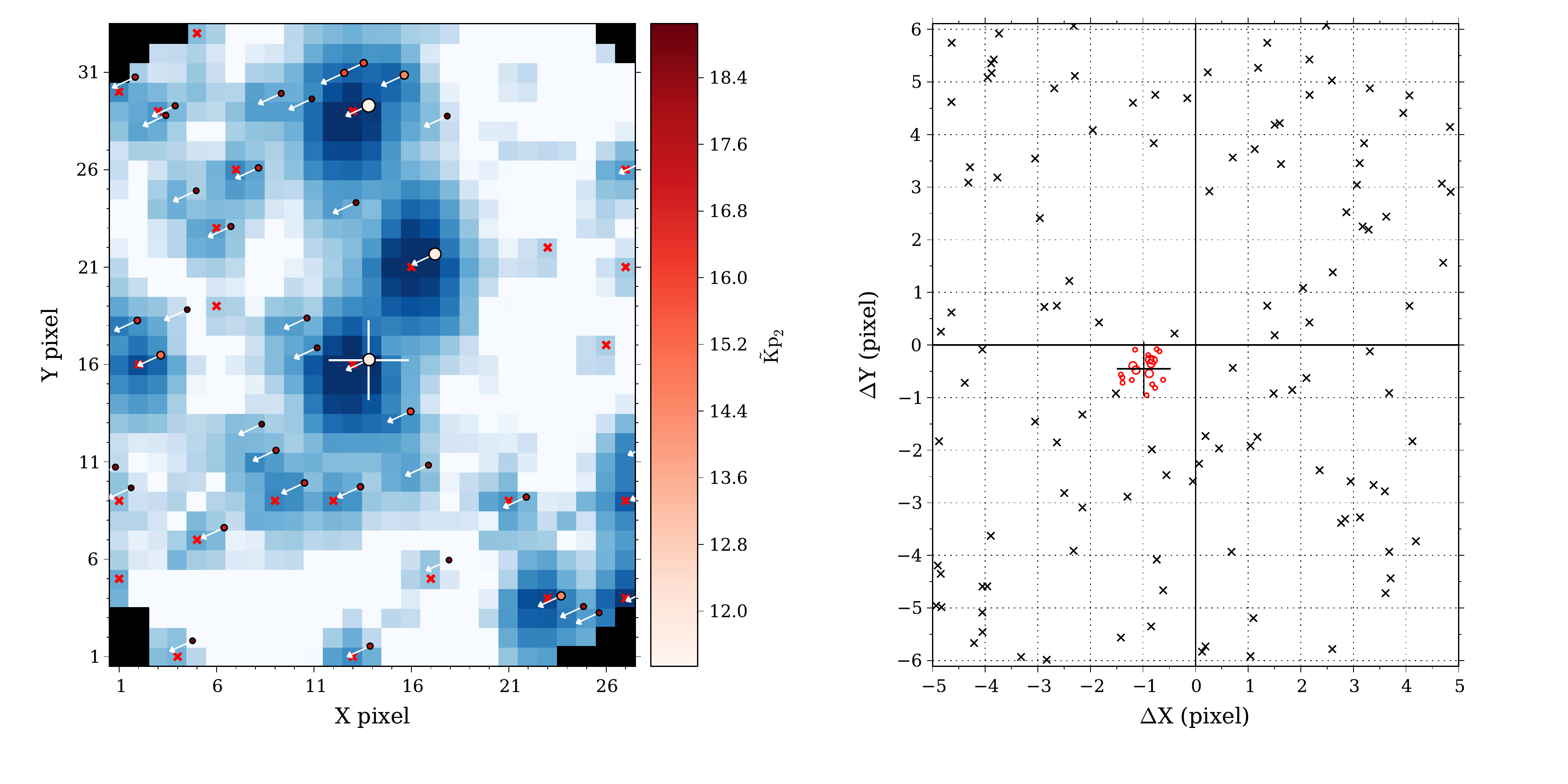}\vspace{-2em}
\caption{Correction of the WCS pixel positions. Left: Circles indicate positions of the targets from the USNO-B1.0 catalogue falling in the frame of EPIC 202127012, where the transformation from sky to pixel coordinates was made using the WCS metric from the K2 pixel target files; target magnitudes \Kptt (see \eqref{eq:mag}) are indicated on the color scale and by the marker size. These are plotted on top of the summed image from C0 data, with a flux scale going from white (low flux) to dark blue (high flux) and rendered on a logarithmic scale. The arrows give the estimated correction to the WCS pixel positions; red crosses indicate the identified maxima in the summed image that are used to estimate the correction. The primary target, \ie, EPIC 202127012, is close to the center of the frame and emphasised with a white ``+'' at the corresponding target position. Right: Position differences between USNO-B1.0 targets and maxima in the summed image. Black crosses are differences tagged as noise in a DBSCAN run on the differences; red circles give the differences of the identified cluster. The final magnitude weighted average of the difference cluster is given by the black ``+''.}
\label{fig:wcs}
\end{figure*}

In K2 a standardised mask is no longer delivered for the main target, at least not in the data releases so far. This, combined with the increased crowding in the equatorial pointing and larger frames, makes it more difficult to assert which target is the main target. Also, the primary target is sometimes fainter than secondary targets in the frame. A starting point for locating the primary target is the assumption that it is (approximately) centred in the frame, but still it will be difficult to use this exclusively in crowded fields. The target pixel files from K2 do deliver a world coordinate system \citep[WCS;][]{2002A&A...395.1061G,2002A&A...395.1077C,2006A&A...446..747G} metric in the FITS format. The WCS from K2 data release 2 is fairly well calibrated (not available in the engineering data) as shown in \fref{fig:wcs}. Here we have marked the positions of all targets from the USNO-B1.0 catalogue \citep[][]{2003AJ....125..984M} from using the WCS transformation to pixel coordinates; it is clear that the WCS delivers a reasonable transformation, generally within two pixels of maxima in the summed images.
An advantage of our pipeline is that masks are defined for all targets in the field (unless they are too faint), and so identification can be made at a later stage. 

So far we have made identifications using the \texttt{Python} module \texttt{Astroquery} \citep[][]{Ginsburg2013}, together with the WCS module in the \texttt{Kaptayn} package\footnote{\url{http://www.astro.rug.nl/software/kapteyn/}}. This enables us to link targets with objects from the USNO-B1.0 catalogue.
The procedure for the identification of targets is as follows: (1) load sky coordinates for all targets located within a circular region which fully contains the frame for the EPIC target in question; (2) Transform the sky coordinates of these targets to pixel positions using the WCS from the K2 target pixel file; (3) Compute a proxy for the \kp magnitude, \Kptt (see \sref{sec:magni} above); (4) Locate maxima in the summed image (\sref{sec:sumimg}) where a $0.5$ pixels wide Gaussian smoothing has been applied; (5) Compute all $X$- and $Y$-pixel differences between the targets and the maxima of the summed image; (6) Run a DBSCAN clustering on all pixel differences within $\pm 5$ pixels in both the $X$- and $Y$-direction, with the clustering parameters set to $r_c=0.5$ pixels and $N_c\leq N_{\rm maxia}$. The value of $N_c$ will initially be the number of identified local maxima and will iteratively be decreased until a cluster is identified in the differences (a ``difference cluster''); (7) If more than one difference cluster is found within $\pm 5$ pixels, we choose the cluster with the lowest mean \Kptt magnitude; (8) As the correction that should be applied to the WCS transformation we take the weighted average of $X$- and $Y$-differences in the difference cluster, using one over the \Kptt magnitudes as weights (see \fref{fig:wcs}). Note that this correction only includes translation, but ignores rotation. This could be amended by using a pattern matching algorithm \citep[see, \eg,][]{a2010093}, but the offsets are low enough that this can be safely omitted; (9) The target from USNO-B1.0 with corrected pixel coordinates closest to the median centroids of identified target clusters is used to identify the cluster. Here we also note if other targets fall in the mask of a given target cluster.


\subsection{Target flux and position}
\label{sec:pos_flux}

The above steps were concerned with the creation of pixel masks for all the different targets in a given frame.
For all these targets we compute the position and flux as a function of time --- this will be used later for the correction of the spacecraft roll. We use weights, $w_{\rm i}$, on the pixels in the individual masks when extracting fluxes and calculating the target position via the centroid (CEN) with ($X$,$Y$) components given as
\begin{equation}
{\rm CEN}_X = \frac{\sum_i{p_{\rm i} w_{\rm i} X_{\rm i}}}{\sum_i{p_{\rm i} w_{\rm i}}  }\, ,\qquad {\rm CEN}_Y = \frac{\sum_i{p_{\rm i} w_{\rm i} Y_{\rm i}}}{\sum_i{p_{\rm i} w_{\rm i}}  }.
\end{equation} 
Here $p_{\rm i}$ denotes the flux for the $i$th pixel in a given mask, and $X_{\rm i}$ and $Y_{\rm i}$ denote the coordinates of the pixel.

We have defined the following three pixel weightings $w_1-w_3$: in $w_1$ all weights are set to $w_{\rm i} = 1$, giving an ``in/out'' mask where all pixels have equal weight; in $w_2$ weights are given as the exact euclidean distance between a pixel in the mask $(X_{\rm i}, Y_{\rm i})$ and the closest background pixel $(X_{\rm i,b}, Y_{\rm i,b})$:
\begin{equation}
w_{\rm i} = \sqrt{(X_{\rm i,b} - X_{\rm i})^2 +  (Y_{\rm i,b} - Y_{\rm i})^2 }\, .
\label{eq:wei}
\end{equation}
In $w_3$ weights are set to create a soft edge on the mask, with a uniformly weighted central region. This is accomplished by dividing every pixel into $11\times11$ subpixels, each of which is assigned a weight given by \eqref{eq:wei} and normalized by $11.3$. If this normalized weight is above 1 it is set to 1. This results in a mask edge of just over $1$ pixels width where the weight gradually increases from one over $11.3$ to 1.

We then tested our pipeline using four different schemes: (1) $w_1$ is used for the extraction of both centroids and flux, here all pixels will influence the position and flux with equal weight; (2) $w_2$ is used for the extraction of both centroids and flux. Such a weighting reduces the sensitivity of the extracted positions to the exact mask configuration, where, for example, a high spatial frequency of the mask from pixels at the mask edges could result in an unwanted flickering in the extracted parameters for a $w_1$ weighted mask. In many ways this resembles the weighting done naturally when using pixel response functions (PRF) to extract centroids \citep[][]{2010ApJ...713L..97B}, but without the need to optimise for centroid and total flux using a parametrised function. The use of \kp calibrated PRFs is further complicated by the fact that the pointing jitter (now with an attitude control bandwidth of $0.02$ Hz until C3, where it will be increased to $0.05$ Hz, which is half of the bandwidth of nominal \kp observation) and systematic movements within a cadence are different in K2 from the nominal \kp mission. Also, for saturated targets the parametrisation fails to represent the flux distribution, and the PRFs are only defined for long-cadence (LC) observations; (3) $w_3$ is used for the extraction of both centroids and flux; (4) $w_2$ is used for the extraction of centroids, while $w_1$ is used for the extraction of flux.

We compared the different weighting schemes in the power spectrum, in the centroids, and in the final corrected time series. No noticeable difference was found between schemes (1), (3), or (4); with this in mind we opt for the simplest scheme, \ie, (1). We also note, first, that scheme (2) gives centroid values with a lower point-to-point scatter; this might be of use for future (from C3) 2D corrections of short-cadence (SC; $\rm \Delta t \approx 1\ min$) data, but seems to have little influence on the 1D correction. Secondly, the flux from scheme (2) retains the greatest signal from the spacecraft movement, which is evident in the power spectra from this method. This might be expected from the peaked flux weighting, which increases the sensitivity to the spacecraft movement --- a flat weighting is preferable for the flux extraction.
Considering the choice between scheme (1) and (3) for extracting the flux, both of which have a predominantly flat weighting, (3) reduces the risk of contamination between targets. Scheme (1), however, makes it easier to identify any contamination that might occur in any case. For simplicity we opt for scheme (1) for both the position and flux extraction, but consider using either scheme (4), or a combination of (3) and (1), for data from future campaigns.

In the final step of extracting the flux from the defined masks we subtract the background level given as the mode of the flux distribution (see \sref{sec:bg}), but now only including pixels that are unassigned to a target mask. If a target is close to the edge, the weighting scheme given by \eqref{eq:wei} will put highest centroid weight towards the edge. However, as long as the flux variability from position correlates with the measured centroid, the data from such edge targets should still be usable. 
The same goes for centroids from saturated targets, as long as the extracted centroids correlate with the relative flux variation they can be used in the correction; the absolute position of the target is of little importance.


\subsection{Contamination between targets}
\label{sec:contam}

Given that most EPIC frames contain multiple targets we compute a few statistics to ascertain the level of contamination between these targets.
As a first metric we compute a contamination value, $C$, as one minus the flux ratio of the primary and all targets in the mask:    
\begin{equation}
C = 1 - \frac{F_{\rm primary}}{F_{\rm total}} = 10^{0.4(m_{\rm total}-m_{\rm primary} )}\, ,
\end{equation}
where $m_{\rm primary}$ is the \Kptt magnitude of the brightest target in the mask; the total apparent magnitude, $m_{\rm total}$, of the mask is given as
\begin{equation}
m_{\rm total} = -2.5\log_{10} \left(\sum_i 10^{-0.4m_i} \right)\, ,
\end{equation}
where $i$ runs over the number of identified stars falling within the given mask. 
\begin{figure}
\centering
\includegraphics[scale=0.25]{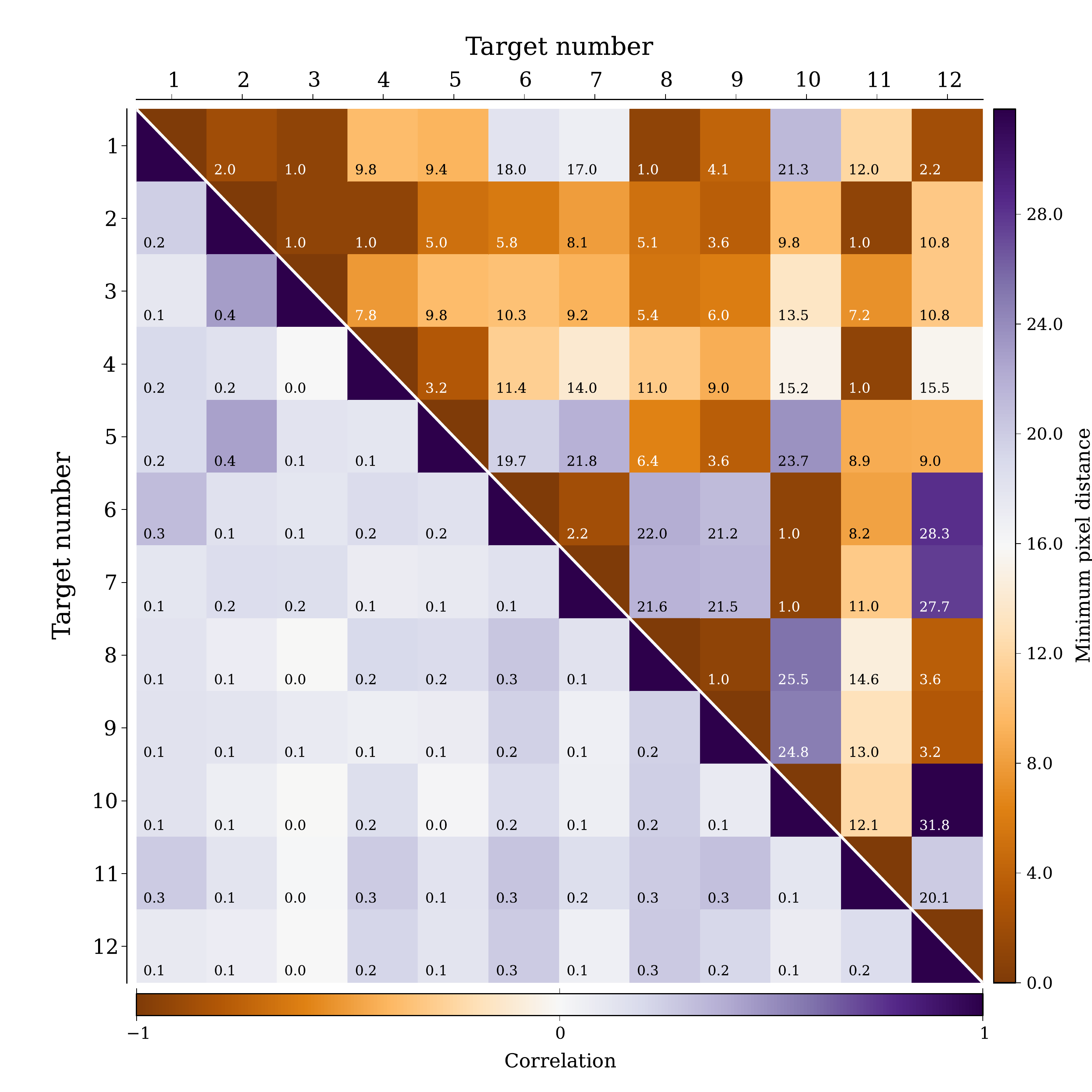}
\caption{Example of a target correlation matrix. The lower left half of the matrix gives the correlation (Pearson's) between power spectra of targets identified in a given frame (see bottom color bar); the top right half gives the corresponding minimum distances between target masks (see right color bar).}
\label{fig:correlation}
\end{figure}
Secondly, for a given frame we compute a target correlation matrix. The lower left half of \fref{fig:correlation} shows the correlation between the targets' power spectra (of the cleaned time series, see \sref{sec:c0ana}); the top right half gives the minimum distance between pixels belonging to each target pair.
This correlation matrix can be used to easily ascertain the contamination between targets, and thus when extra care should be exercised in assigning a given signal to a given star. 


\section{Correcting the light curve}
\label{sec:corr}

We have combined the correction part of our pipeline with the KASOC filter \citep[][]{2014MNRAS.445.2698H}, meaning that the corrections based primarily on the target movement on the CCD are combined with corrections made for long and short term instrumental trends via the KASOC filter --- and this in an iterative manner. 
Briefly, the KASOC filter works by computing two median-filtered versions
of the time series with different filter windows, and then forms a weighted combination of the two to correct
the time series for instrumental features. We refer to \citet[][]{2014MNRAS.445.2698H} for further details on the KASOC filter.
The integration with the KASOC filter also includes the iterative use of phase curve corrections, which is particularly useful for separating the flux variations from the target movement on the CCD from those of stellar variability with a strict periodicity (for instance the eclipses of a planetary or binary system).

Below we describe the two possible correction methods in the pipeline. For both methods it generally holds true that when the amplitude of the underlying stellar signal dominates the variations, such as in many Classical pulsators, the correction of the instrumental signal is less effective. 


\subsection{1D correction}
\label{sec:corr1}

Our 1D correction draws heavily on the method presented by \citet[][]{2014PASP..126..948V} --- which these authors called a self-flat-fielding correction --- which in turn make some use of methods developed for correction of \emph{Spitzer} data \citep[][]{2008ApJ...673..526K,2010PASP..122.1341B,2012ApJ...754..136S}.
These methods use the correlation between flux variation and position on the CCD (from pixel sensitivity differences across the CCD), to correct the time series from the systematic ${\sim}6$ hour variability.   

\begin{figure}
\centering
\includegraphics[scale=0.4]{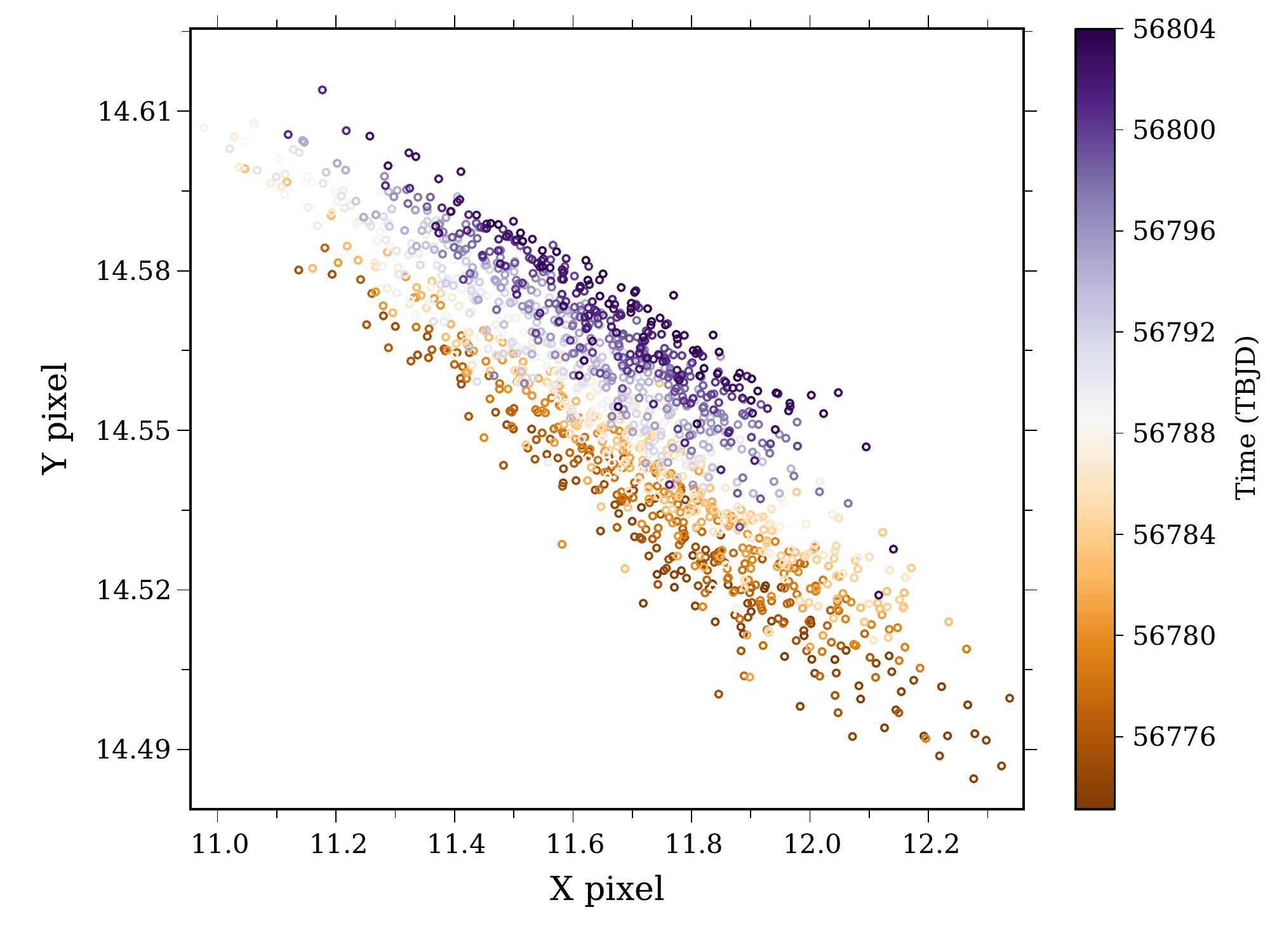}
\caption{Illustration of the change in centroid position of star 1 in EPIC 202127012 on the CCD during the second half (approximately) of C0; time is encoded in the colour scale. A sigma clipping has been applied in the time domain to remove point far away from the mean centroid position.} 
\label{fig:2dpos_vs_time}
\end{figure}

We break the time series into segments that are corrected individually. This segmentation was implemented because even though the movements on the CCD generally follow a well-defined pattern (which depend on position on the focal plane), there are slow uncorrected drifts as a function of time (see \fref{fig:2dpos_vs_time} for an example of this in C0). Currently, the times where breaks are introduced are determined manually, and are kept constant for all targets in a given campaign; we provide flags for the times where breaks are introduced in the final output. For C0 the time series was broken into two segments, namely, a ${\sim}13$ day segment before and a ${\sim}35$ day segment after a safe mode event occurring in C0 (lasting approximately $24$ days). 

For each segment, we start by identifying and flagging times during which a rapid positional change occurs, as the times when the time derivative of the change in centroid positions, \ie, the velocity, falls outside the range of five times the standardised MAD\footnote{which we define as $1.4826$ times the MAD, with the constant being the scale factor that makes the MAD a consistent estimator for the standard deviation of a normally distributed random variable.} around the median velocity; these data points are then excluded in the following corrections. 

We then apply a principle component analysis (PCA) on the $X$ and $Y$ pixel positions of each data segment. Before applying the PCA, we select which of the $X$ and $Y$ pixel positions should be retained in the estimate of the correction; only positions with a nearest neighbour at a distance less than four times the standardised MAD of all nearest neighbour distances are retained in the analysis. This is needed as the PCA otherwise is very sensitive to outliers. The PCA transformation of the retained positions to the coordinate system given by the two first principal components, helps to ensure that the relationship between the transformed pixel positions $X'$ and $Y'$ can be described as a single-valued function, which is needed for the following steps in the correction.
It is, however, not always clear if the first or the second principal component should be used as the regressor. If, for instance, the relationship between $X$ and $Y$ pixel positions could be described as $Y = X^2$ (which is already a single-valued relationship) and the range in $Y$ values is larger than the range in $X$ values, then the first principal component would lie along the ordinate and consequently a transformation making this the regressor, that is, $Y\rightarrow X'$ and $X\rightarrow Y'$, would result in the multi-valued relationship $X' = \pm \sqrt{Y'}$. We decide which of the principal components is the best regressor, by running a LOWESS \citep[][]{lowess,MR556476} filter on the transformed pixel positions, using in turn the two principal components as regressors and computing the summed squared difference ($\chi^2$) between the filtered and un-filtered data. The principal component with the lowest $\chi^2$ is used as the regressor.

In the transformed coordinates we compute a smoothed version of the $Y'$ vs. $X'$ positions by again applying a LOWESS filter. We then calculate the curve length $s$ along this filtered relationship as
\begin{equation}
s = \mathlarger{\int}_{X'_0}^{X'_1}{\displaystyle \sqrt{1 + \left(\frac{dY'_{\rm LOWESS}}{dX'} \right)^2 } dX'}\, ,
\end{equation}
using finite differences as the derivative of the curve and cumulatively integrate for the curve length using the composite trapezoidal rule. The curve length serves as the new 1D representation of the 2D stellar position on the CCD. 

\begin{figure}
\centering
\includegraphics[scale=0.4]{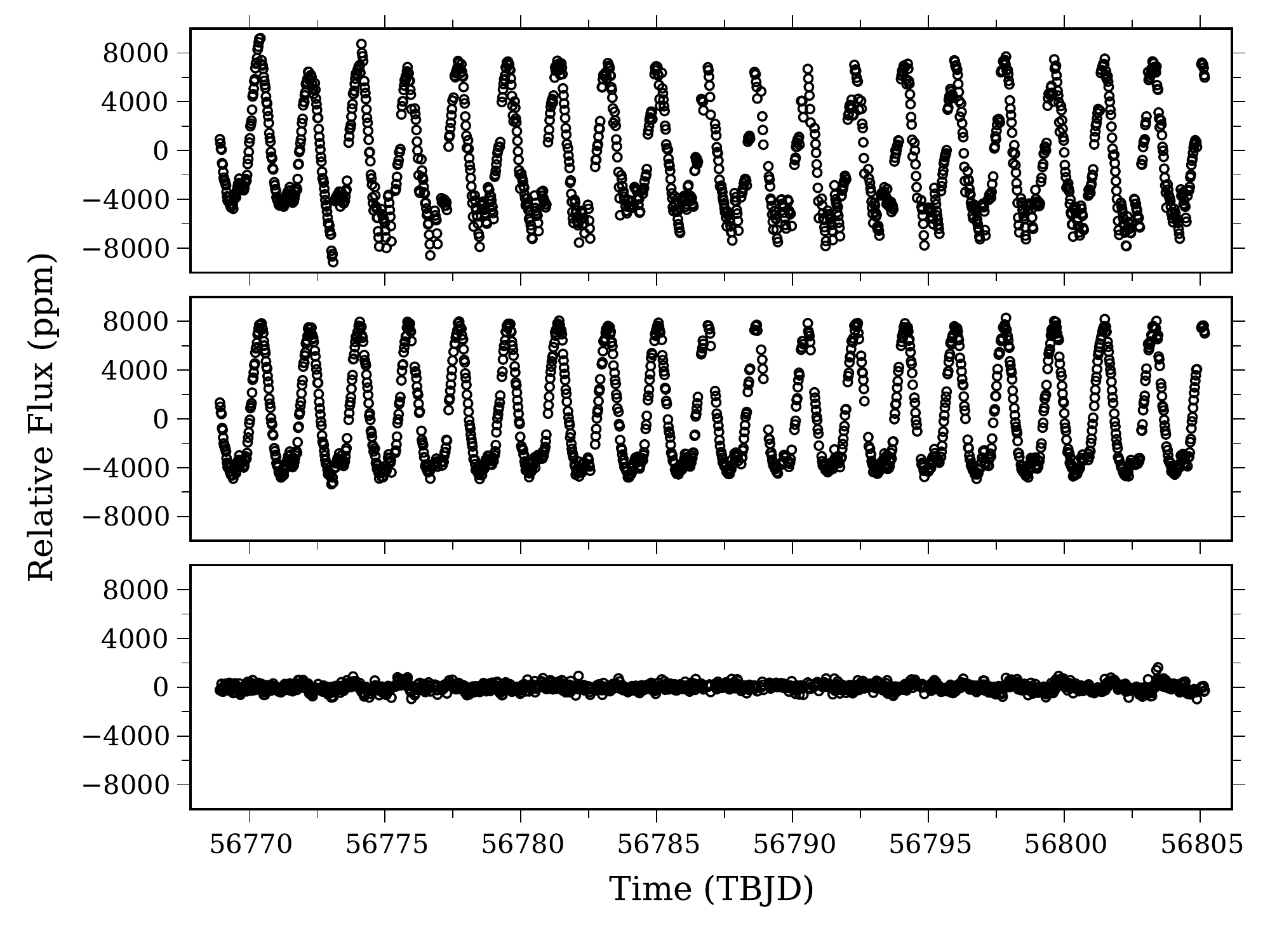}
\caption{Illustration of 1D correction for star 2 (TYC 1329-1325-16) in EPIC 202127012 (see \fref{fig:pro2}). Top: Uncorrected time series in flux relative to the median. Middle: Time series corrected for long term, short term, and positional trends. Bottom: Time series corrected for long term, short term, positional trends, and the phase curve constructed from the middle panel (see \fref{fig:1dcorr_phase}).} 
\label{fig:1dcorr_second}
\end{figure}
\begin{figure}
\centering
\includegraphics[scale=0.4]{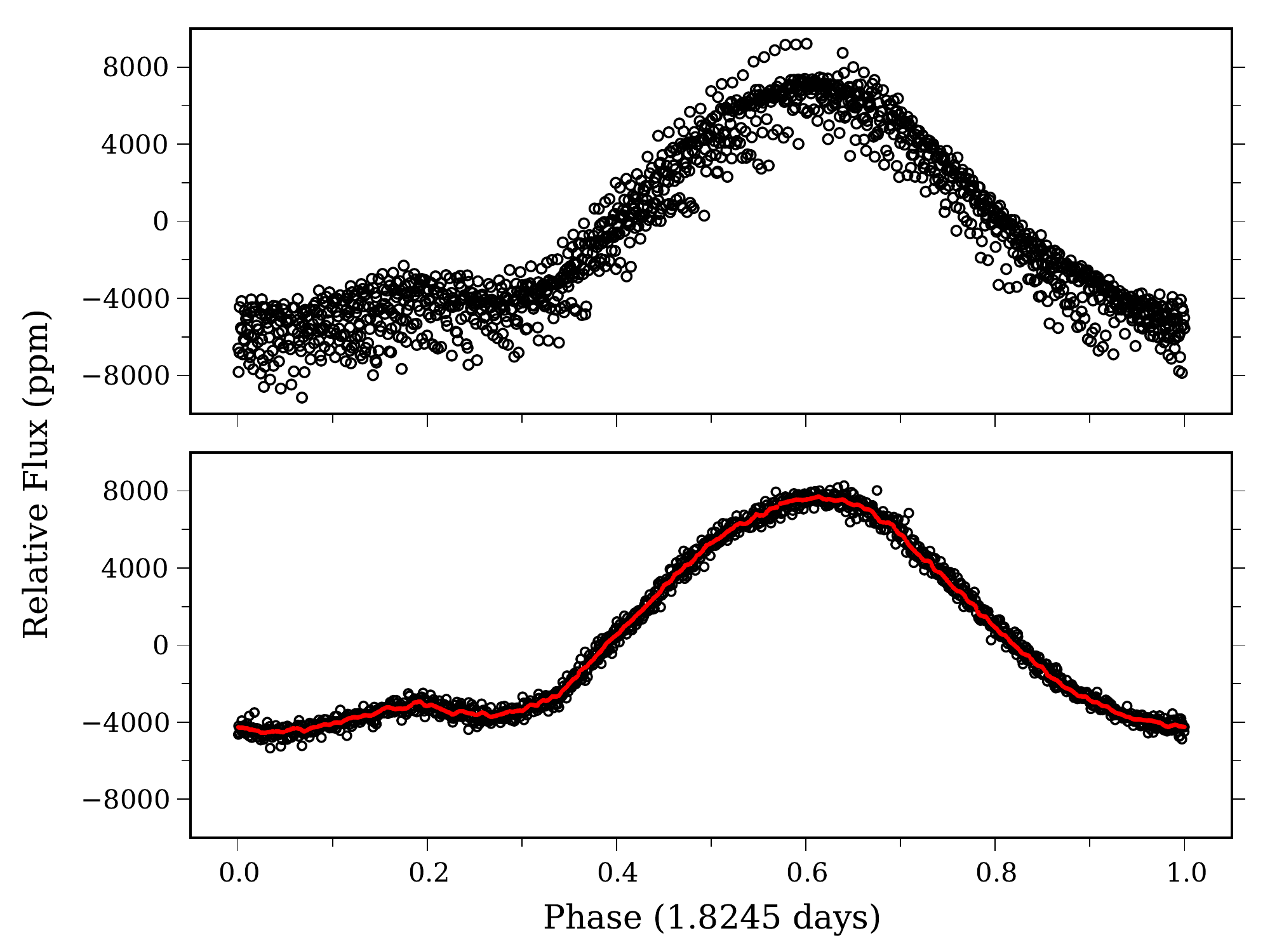}
\caption{Phase curve for star 2 (TYC 1329-1325-16) in EPIC 202127012 (see \fref{fig:pro2}). Top: Phase curve of uncorrected time series with the flux relative to the median (top panel in \fref{fig:1dcorr_second}). Bottom: Phase curve (black) of time series corrected for long term, short term, and positional trends (middle panel in \fref{fig:1dcorr_second}). From the black points we form the red phase curve via a moving median smooth; this smoothed phase curve is used in the iterative correction performed by the KASOC filter to obtain the bottom panel of \fref{fig:1dcorr_second} \citep[][]{2014MNRAS.445.2698H}.} 
\label{fig:1dcorr_phase}
\end{figure}
The correction to the light curve is then found from a LOWESS filtering of the relative flux as a function of curve length, thereby capturing the average positional dependency of the flux level. 
In the correction step we make sure to remove any long term trends in the light curve to obtain the relative flux, as such changes will correlate poorly with the movement on the CCD. Some of the long term variability could in principle be caused by the slow drift of the target on the CCD (\fref{fig:2dpos_vs_time}), but could just as well be a separate instrumental effect --- for instance from focus changes caused by heating of the mirror. The background flux level could also enter in the long term variability if this is not corrected for properly during the light curve extraction. We make the correction iteratively with a better separation between long term and positional dependent variations as the outcome. 

In \fref{fig:1dcorr_second} we give an example of the 1D correction for the C0 observations of star 2 (TYC 1329-1325-16) in EPIC 202127012 (see \fref{fig:pro2}); here we further include in the KASOC filter a correction for the dominating periodic signal by iteratively correcting by the phase curve of this signal (see \fref{fig:1dcorr_phase}). The input period for this correction was determined from the autocorrelation function of the time series.


\subsection{2D correction}
\label{sec:corr2}

\begin{figure}
\centering
\includegraphics[scale=0.45]{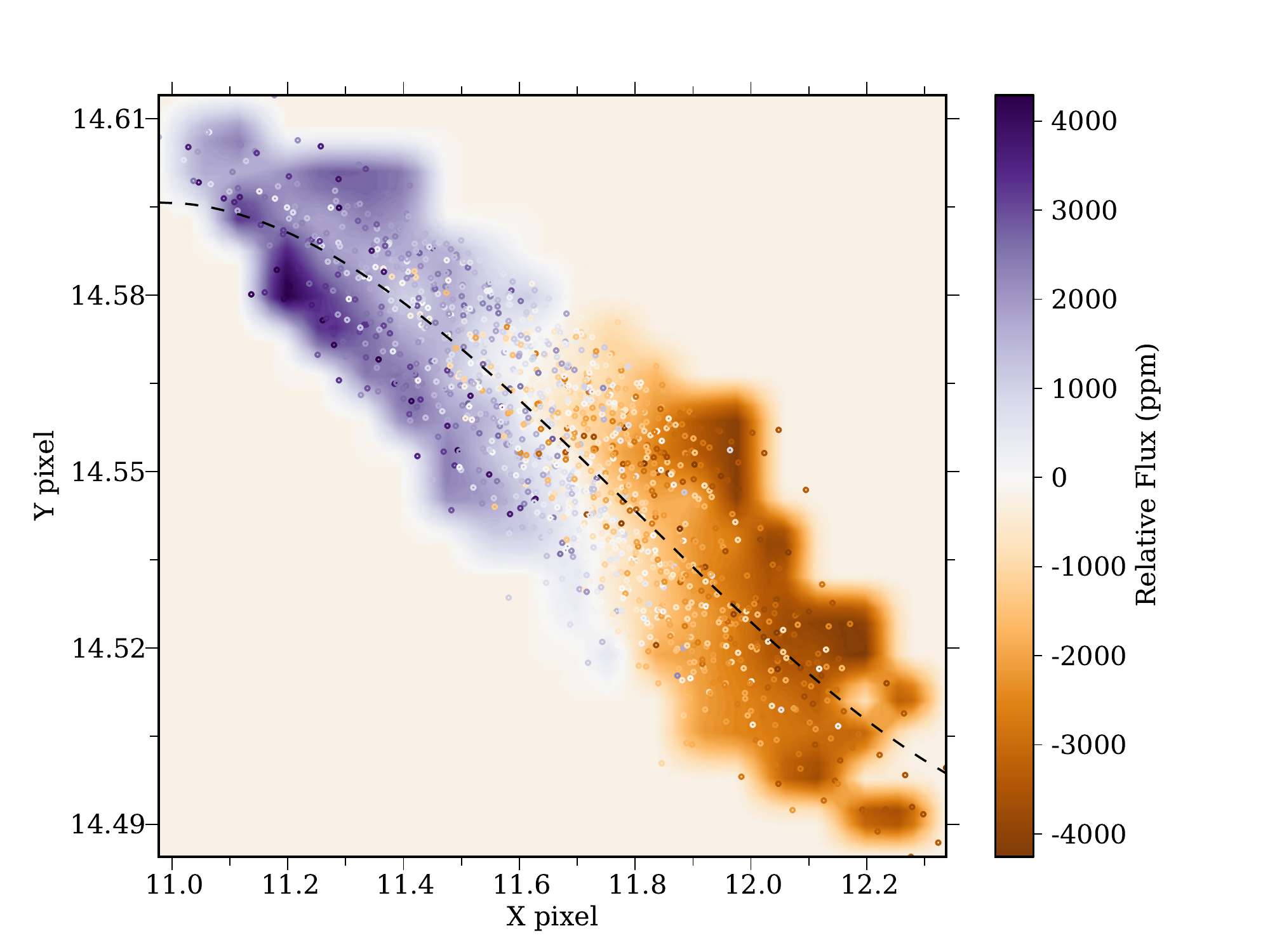}
\caption{Centroid positions for star 1 in EPIC 202127012, with the relative flux given by the colour bar. The surface shows the interpolated relative flux from the medians of the 2D histogram. Here we used 20 bins in both the X and Y-directions. The dashed line gives a smooth version of the overall positional variation, that could be used to correct the positions to a more one dimensional variability.} 
\label{fig:2d_corr}
\end{figure}

In our second approach we make a 2D histogram of the measured X and Y centroids of the star. In each bin we compute the median of the relative flux of points falling in that bin; this will capture the positional variation in the relative flux in a robust manner. In the reconstruction of the flux variability in the time domain we use a rectangular bivariate linear spline to interpolate between the bin centres. The reason for going to 2D is that flux variations also occur in the direction perpendicular to the overall roll motion (see \fref{fig:2d_corr} for an example). Such variations are unresolved in the 1D treatment, because the scatter in the relative flux versus curve length is reduced to a line; one would therefore suspect that the 1D treatment will leave residuals in the corrected light curve that could be accounted for in a 2D treatment.

The most difficult aspect of the 2D binning is the choice of bin size. If the bins are too small the reconstruction of the flux variation will be noisy; one is effectively overfitting. On the other hand, if the bins are too large the reconstructed variation will be a smoothed version of the underlying variation, and significant residuals may be left in the light curve. The sensitivity to the bin size is largest for long-cadence (LC; $\rm \Delta t \approx 29.4$ minutes) observations due to the smaller number of data points, and consequently larger variance on the median. The method is thus best suited for SC observations where the exact bin size is less influential on the reconstructed instrumental variability.

Depending on the shape of the stellar movement in the X-Y plane it can be advantageous to transform the movements to a (predominantly) horizontal variation before making the 2D histogram --- this could, for instance, be achieved by dividing the centroid Y-components with a smoothed fit to the movement in order to reduce the span of the histogram, and thus the size, of the histogram.

So far, in our testing of this method on SC data we have not found it to be preferable to the 1D method. This is likely a result of the current value of the attitude control bandwidth of $0.02$ Hz ($50$ s), which is very close to the SC integration time. Because of the allowed amount of movement within a SC integration this will lead to a larger smear and variance in the bin medians. We expect this to improve from C3 onwards when the bandwidth will be increased to $0.05$ Hz.


\section{Pipeline test}
\label{sec:c0ana}

As a test of the pipeline we analysed the pixel frames of the 452 LC targets in the C0 proposal GO0118\footnote{\url{http://keplerscience.arc.nasa.gov/K2/docs/Campaigns/C0/GO0118_Stello.pdf}} (``Galactic Archaeology on a grand scale''; PI: Stello, D.). We also analysed the known transiting system WASP-85 \citep[see][]{2014arXiv1412.7761B}, which was observed in SC during C1.

Because our pipeline enables the extraction of data from several targets in a given frame, we ended with a total of $4691$ targets from the GO0118 proposal, and thus light curves to analyse --- this corresponds to a gain in the amount of data by a factor of ${\sim}10.4$, and this even when adopting a limit on the minimum number of pixels in a mask of $8$ before a target would be considered.   


\subsection{The power spectrum}

\begin{figure*}
        \centering
        \subfigure{
                \includegraphics[scale=0.4]{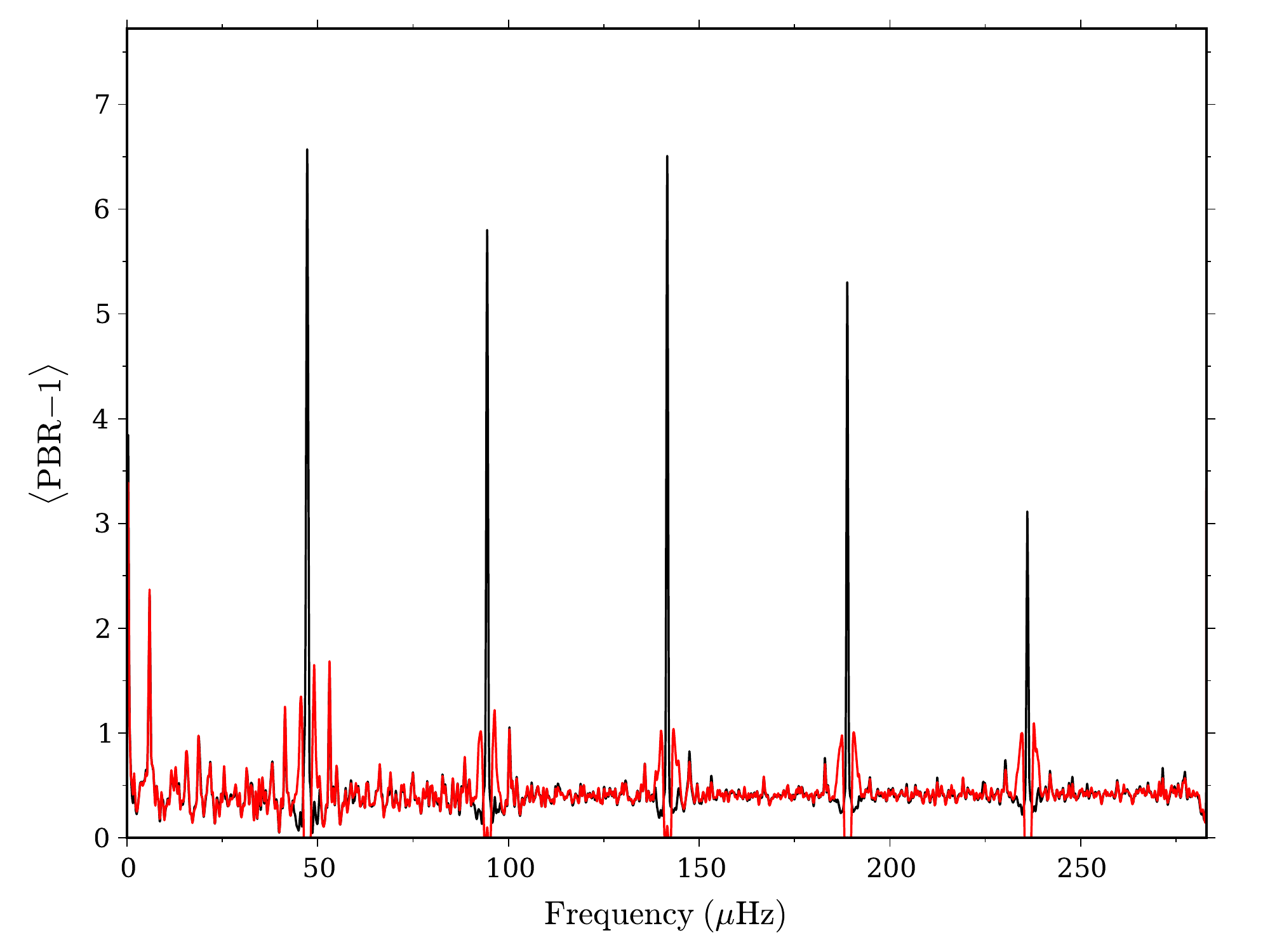}
        }\hspace{1cm}
        \subfigure{
                \includegraphics[scale=0.4]{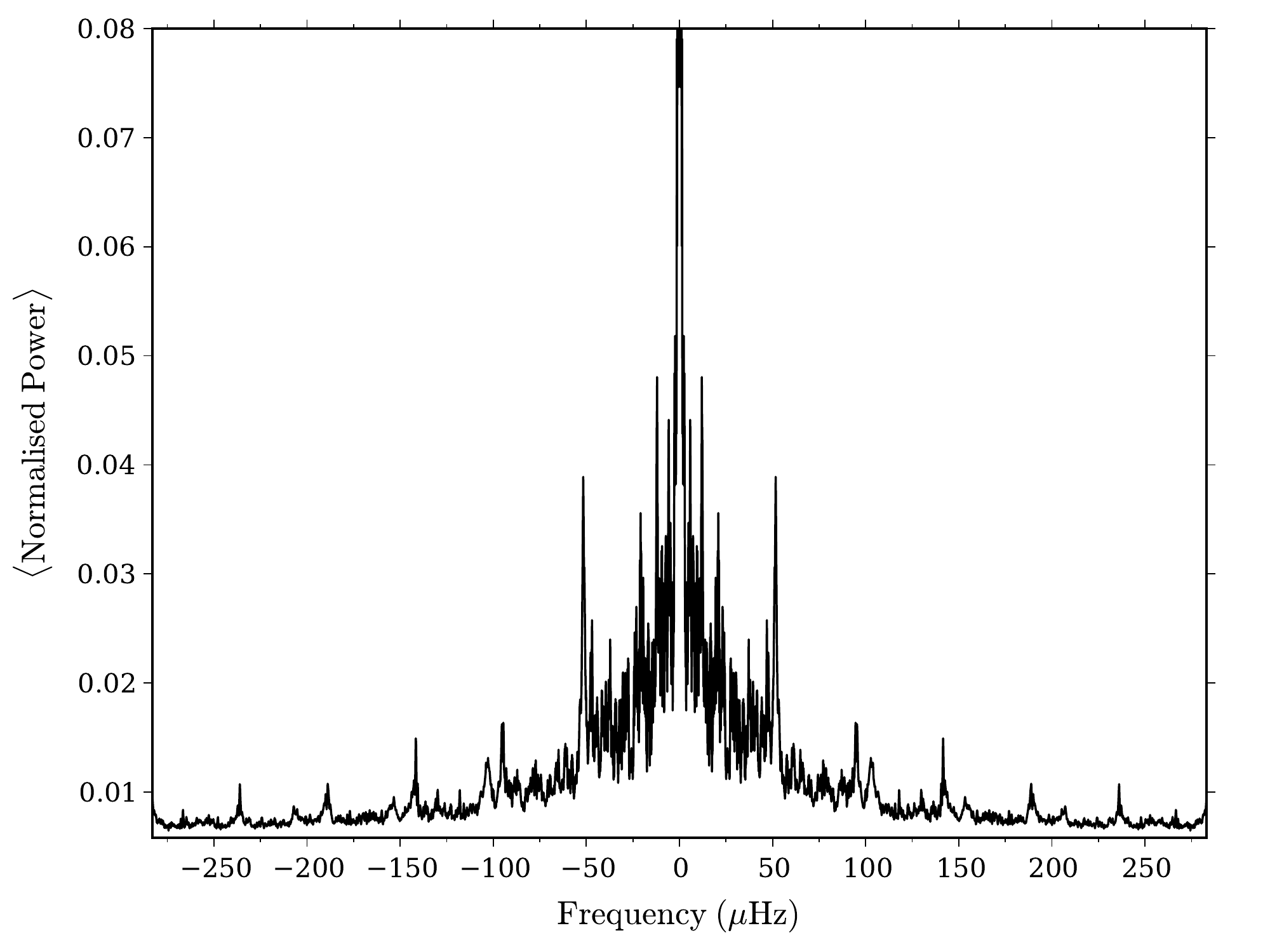}
        }
        \caption{Left: Effect of cleaning residual peaks from the space craft roll. The black curve gives the mean of the 4691 power spectra; each of which have been divided by a $\rm 5\, \mu Hz$ window running median smooth to convert to a power-to-background ratio (PBR). Here the residuals from the space craft roll is clearly visible at integers of $\rm {\sim}47.2281\, \mu Hz$. The red curve gives the spectrum after prewhitening the residual peaks. Right: Average spectral window for the 4691 time series, normalised to 1 at zero frequency. In the averaging of the power spectrum and spectral window both of these were interpolated onto a common frequency scale using a smoothing spline interpolation.}\label{fig:clean}
\end{figure*}

After data were extracted using the K2P$^2$ pipeline, they were corrected with the KASOC pipeline \citep[][]{2014MNRAS.445.2698H} using the 1D correction method described in \sref{sec:corr1}, and a frequency power density spectrum was calculated. The 1D correction removes most of the signal from the spacecraft roll, but residual spikes still often appear at harmonics of $\rm {\sim}47.2281\, \mu Hz$. These spikes are damaging to any automated search for power; to remedy this we tested the effect of ``cleaning'' the residual spikes using a prewhitening routine \citep[see, \eg,][]{1981MNRAS.196..583P,1991A&A...246...71B}, which removes all significant power in a $\rm \pm1\, \mu Hz$ window around the residual spikes. For every window, oversampled by a factor of ten, we iteratively remove the frequency with the highest power-to-background ratio (PBR; the background is calculated as the median of the power within the window multiplied by ${\sim}1.42$, which is the conversion factor between the median and the mean for a $\chi^2_2$-distribution) if this ratio has a false-alarm detection probability less than $10\%$ \citep[][]{1982ApJ...263..835S,2004A&A...428.1039A,2012MNRAS.427.1784L}. Besides the signal from the spacecraft roll, we also see a signal at $\rm {\sim}5.92\, \mu Hz$ (equivalent to $\rm {\sim}1.96$ days); we suspect this signal originates from the periodic momentum dumps of the reaction wheels through thruster firings, which happens every two days \citep[][]{2014PASP..126..398H}, and enters the power spectrum via the spectral window (see right panel in \fref{fig:clean}). The left panel of \fref{fig:clean} shows the efficiency of the procedure for removing the residual instrumental peaks from the power spectrum. Instrumental signals can still be seen in the cleaned power spectrum, but now with amplitudes low enough to allow the detection of asteroseismic signals.  

\subsection{High frequency photometric variability}
\begin{figure}[htp!]
\centering
\includegraphics[scale=0.4]{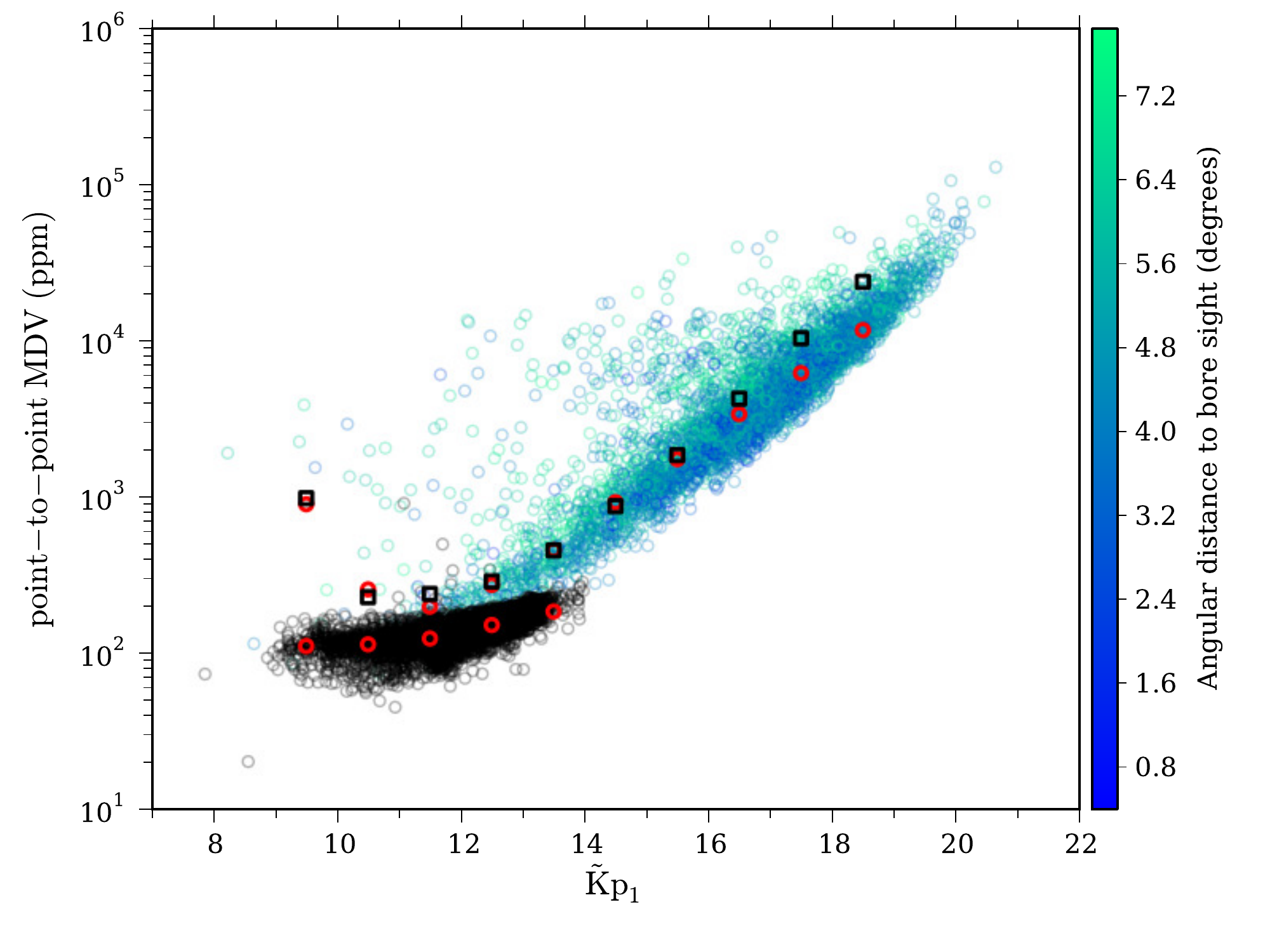}
\caption{Proxy for the short time scale (high frequency) noise, given by the point-to-point median differential variability (MDV), as a function of the proxy \kp magnitude \Kpt. Circular colored markers (blue to green) give the estimates for the K2 sample, with the color scale indicating the angular distance to the space craft bore sight (see color bar); circular black markers give the estimates for APOKASC LC targets from the nominal \kp mission (for these targets their actual \kp magnitudes were used); red circular markers give the median MDVs for both K2 and nominal \kp values in 0.5 magnitude bins; square black markers give the median MDVs for K2 from \citet[][]{2014arXiv1412.6304A}.} 
\label{fig:mag_vs_noise}
\end{figure}
To detect stellar oscillations in the frequency power spectrum it is important that the white (shot) noise level does not dominate the signal --- this is especially true for the detection of low amplitude stochastic solar-like oscillations. It is thus of interest to know the characteristic levels of the short time scale (high frequency) noise in K2 LC data as a function of \kp magnitude, or in our analysis \Kpt. We note, however, that a measure of the high-frequency noise is not necessarily tantamount to a measure of the constant power spectral density white noise level. For each of the targets in the sample we computed a proxy for the instrumental variability using the median of the absolute point-to-point flux difference of the KASOC corrected and cleaned time series; this proxy was coined the median differential variability (MDV) by \citet[][]{2013ApJ...769...37B}. As detailed in \citet[][]{2013ApJ...769...37B} the MDV will on short time scales (with point-to-point being the shortest) be most sensitive to high frequency noise; variability on time scales longer than the LC sampling of ${\sim}29.4$ minutes will on the other hand contribute very little to the MDV. To enable a comparison of the MDV for K2 with that of the nominal \kp data, we compute the point-to-point MDV for the set of 6210 LC targets from the \kp APOKASC \citep[][]{2014ApJS..215...19P} data release 1 sample.
In the KASOC filtering we used the following filter settings: $\tau_{\rm long}=3\ \rm days$ and $\tau_{\rm short}=0.25\ \rm days$ (see \citet[][]{2014MNRAS.445.2698H} for details on these settings); for the APOKASC targets we used $\tau_{\rm long}=30\ \rm days$, which is a too long time scale for the duration of the K2 light curves. \fref{fig:mag_vs_noise} shows the resulting MDV measures as a function of magnitude for both the K2 and nominal \kp targets. Our results from the nominal \kp data are in overall agreement with the results presented in \citet[][]{2013ApJ...769...37B}. We find that at $\Kpt\lesssim 10$ the ratio between the median MDV in K2 and nominal \kp falls below ${\sim}2$, and increases to ${\sim}10$ at $\Kpt\sim 14$. For the K2 values we further see an indication of a slight gradient in the MDV with angular distance to the bore sight for a given magnitude, which might be expected from the larger systematic imprint on the light curve further away from the bore sight. Comparing our values to those from \citet[][]{2014arXiv1412.6304A} (their Table~1, 3-pixel radius masks) we find, as evident from \fref{fig:mag_vs_noise}, an excellent agreement. We also computed point-to-point MDVs for our target sample as corrected in \citet[][]{2014arXiv1412.1827V}\footnote{\url{https://www.cfa.harvard.edu/~avanderb/k2.html}}, and find that the median binned values generally agree within a factor of two. For these comparisons it should be noted that we are unaware if the authors of the comparison studies checked the sources of the \kp magnitudes from the TPD, entering the magnitude calibration, and how they possibly transformed these.

We note that a comparison of MDVs can not be seen directly as a comparison of the quality of the light curves and the corrections applied, and should be evaluated in the context for which the corrected data is intended. A measure like the MDV will depend strongly on the choice of free parameters in the correction. In \citet[][]{2014arXiv1412.1827V} the C0 light curves were processed with the intent of detecting planets. Here the light curves are corrected individually in three segments; the values from the mask with the lowest 6-hour scatter were adopted, trying 20 masks of different sizes; the fit to the flux versus curve length was made with a finer binning than in \citet[][]{2014PASP..126..948V} --- all of these tweaks will conspire to giving a lower point-to-point scatter suited for planet detection.


\subsection{Target examples}

In the following we show a few examples of the many targets among the 4691 that display astrophysical signals.
We note that we have not performed a systematic assessment of the targets.

\begin{figure*}
\centering
\includegraphics[scale=0.4]{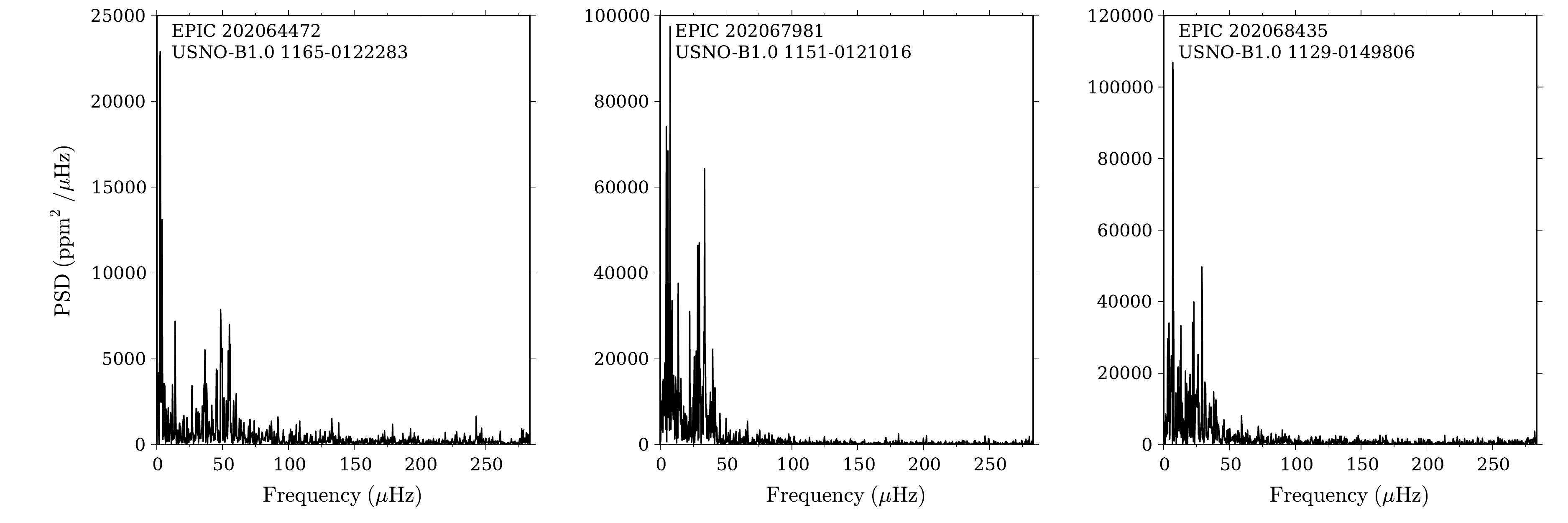}
\caption{Example power spectra of red giant targets, from GO0118 observed during C0, showing low-frequency solar-like oscillations. All of the targets show here are the primary ones for the respective EPIC numbers.} 
\label{fig:power_exam}
\end{figure*}

\fref{fig:power_exam} gives an example of three red giant targets, showing low-frequency solar-like oscillations. The levels of power here suggest that for C0 it should in general be possible to detect oscillations in red giants, and obtain average asteroseismic measures such as $\Delta\nu$ and \numax. We note that for the three cases show in the \fref{fig:power_exam} the \Kpt magnitudes were all ${<}11$, and the high frequency noise in the time domain (approximated by the MDV) is, according to \fref{fig:mag_vs_noise}, only about 2-3 times higher in K2 compared to the nominal \kp mission. If we assume the MDV scales linearly with the shot noise, this translates to a factor of 4-9 times higher noise in the power density spectrum compared to the nominal \kp mission. For a systematic analysis of the C0 red giants we refer to Stello et al. (in prep.).

\begin{figure*}
\centering
\includegraphics[scale=0.4]{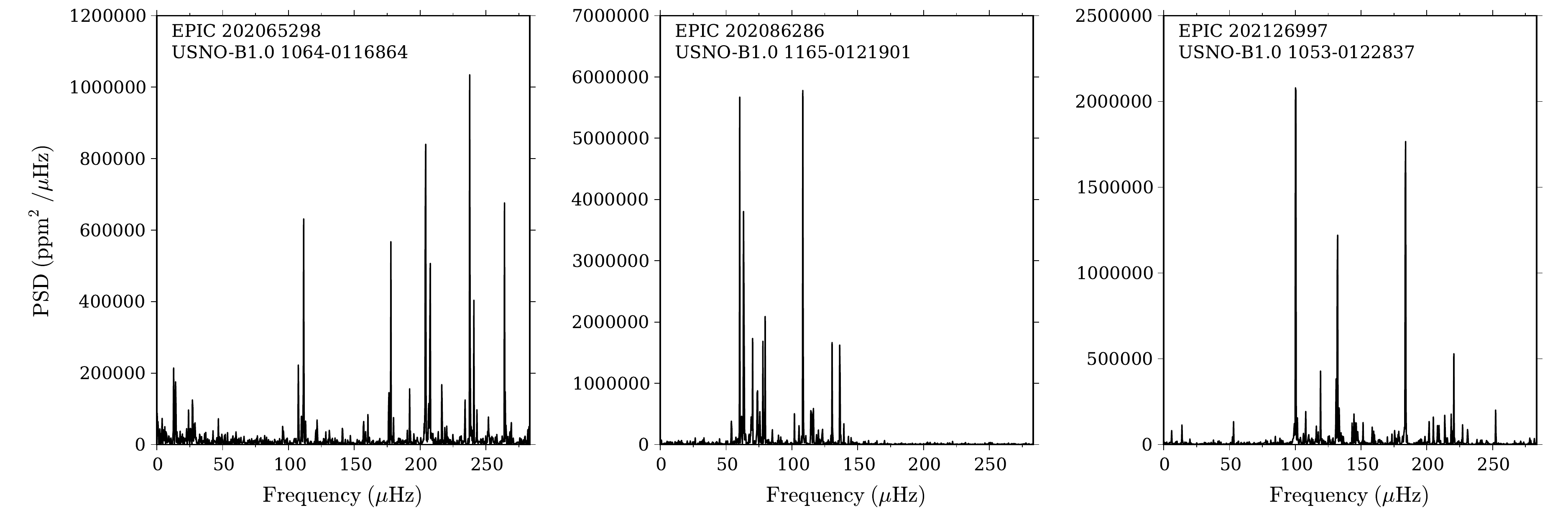}
\caption{Example power spectra of targets, from GO0118 observed during C0, showing classical $\delta$-Scuti-like oscillations, and possibly some $\gamma$ Dor oscillations. Only in the case of EPIC 202086286 (middle panel) does the target correspond to the primary target.} 
\label{fig:power_exam2}
\end{figure*}

\fref{fig:power_exam2} gives an example of three Classical pulsators showing, predominantly, $\delta$-Scuti-like oscillations. For this type of star the noise introduced in K2 is clearly of little importance due to the large amplitudes of the oscillations. 

\begin{figure}
\centering
\includegraphics[scale=0.4]{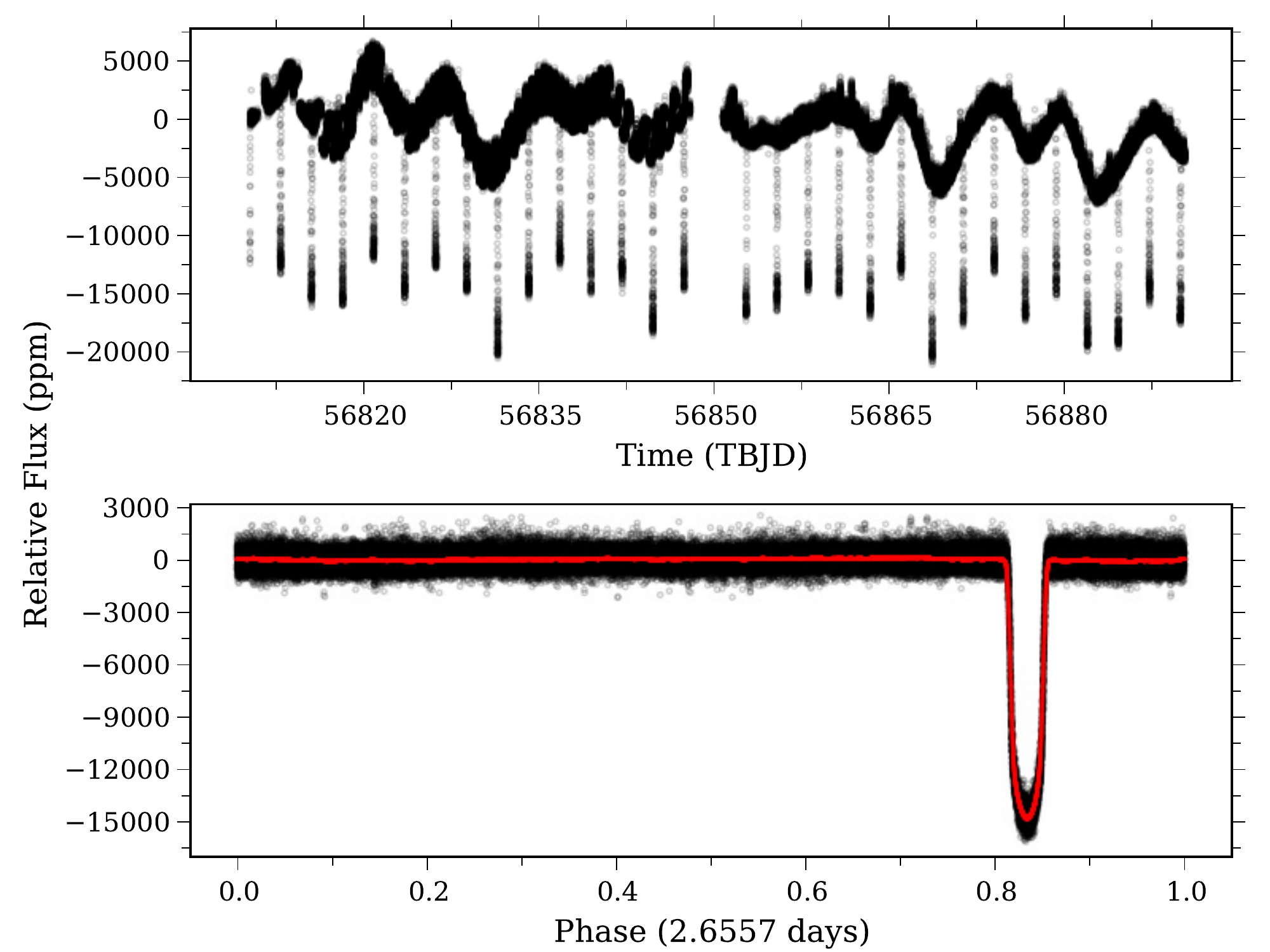}
\caption{SC data for WASP-85 obtained during C1. Top: Raw light curve for WASP-85 showing spot modulation and distinctive transits. Bottom: Phase curve of the corrected light curve for WASP-85; the red curve gives the smoothed phase curve that would have been used in the final correction of the light curve.} 
\label{fig:wasp85}
\end{figure}

In \fref{fig:wasp85} we present the SC data and corrected phase curve for WASP-85 \citep[][]{2014arXiv1412.7761B}, having the EPIC number $201862715$. The raw data for this system shows a clear modulation from surface spots, together with the smaller amplitude instrumental modulation. In the reduction of this light curve we used the information of the orbital period of the system in the iterative correction performed by the KASOC filter. The bottom panel of \fref{fig:wasp85} gives the phase curve at the final iterative step. 

\begin{figure*}
\centering
		\subfigure{
                	\centering
                \includegraphics[scale=0.3]{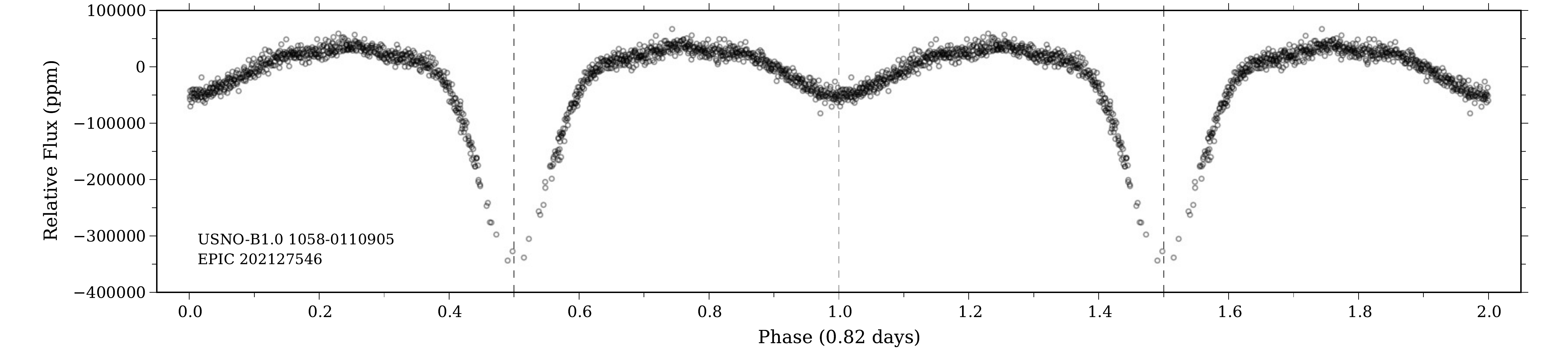}
        }\\ \vspace{2em}
        \subfigure{
                	\centering
                \includegraphics[scale=0.3]{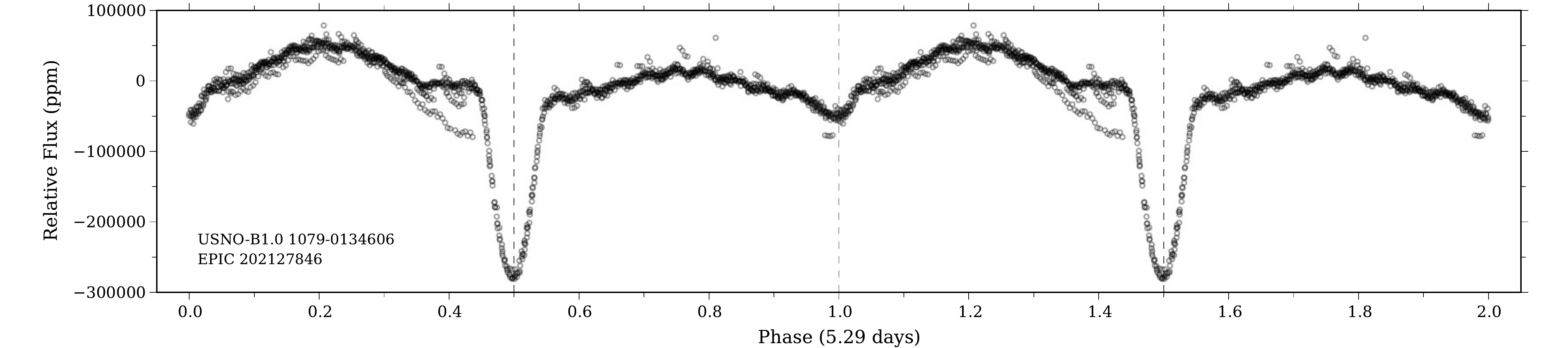}
        }\\ \vspace{2em}
        \subfigure{
                	\centering
                \includegraphics[scale=0.3]{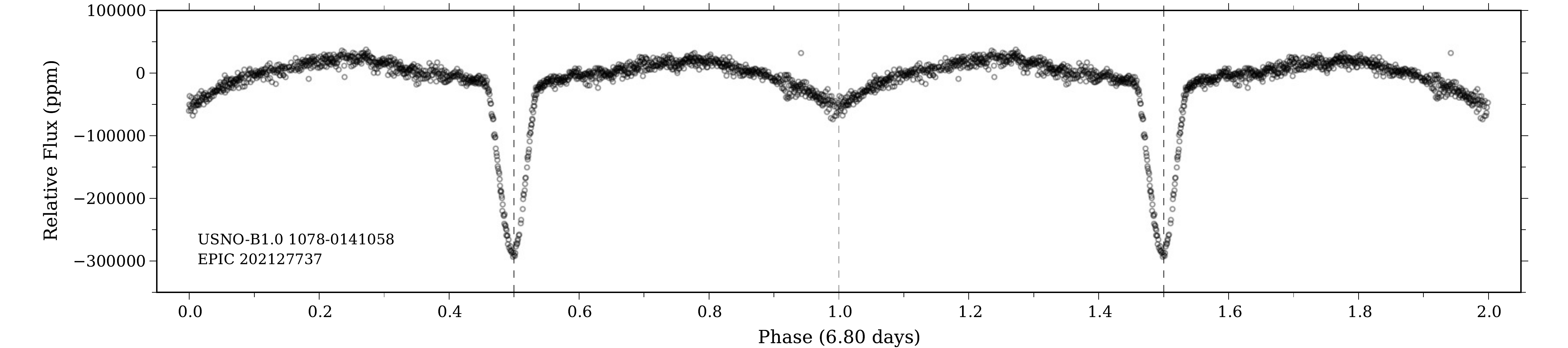}
        }\\ \vspace{2em}
        \subfigure{
                	\centering
                \includegraphics[scale=0.3]{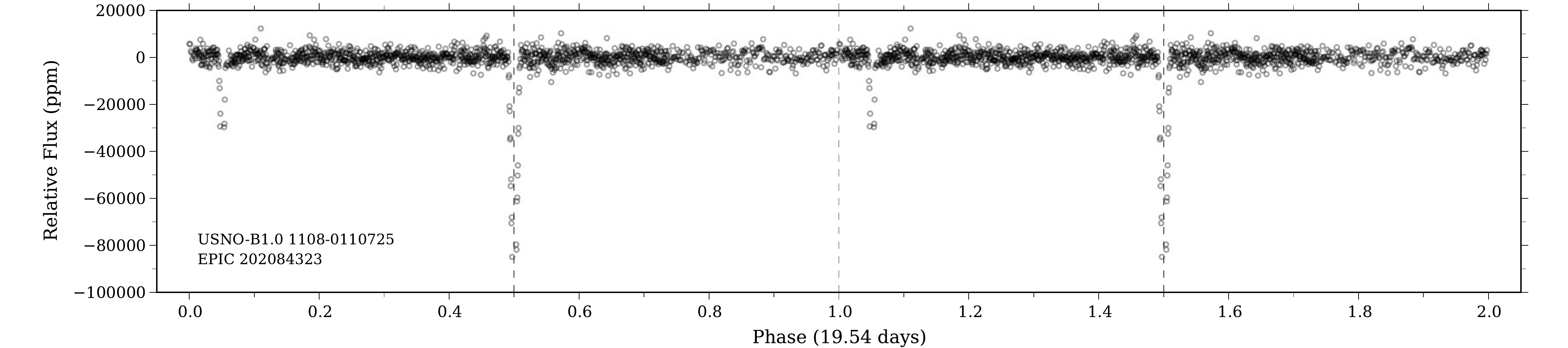}
        }\\ \vspace{2em}
        
        \caption{Phase curves for a subset of targets exhibiting transit-like features, rendered over two full phases with the first primary transit feature centred at 0.5; the approximate period for each target is given in the label of the abscissa. For the purpose of this illustration periods were estimated by eye from the light curves. In each panel we have indicated what we find the most likely USNO-B1.0 number for the target. The EPIC number given in each panel if for the star in whose pixel frame the target was found. In neither of the cases show did the target correspond to the main target associated with the EPIC number.}\label{fig:phaseex}
\end{figure*}

In \fref{fig:phaseex} we present the light curves for a few targets showing distinct eclipse-like features. We note that in none
of these cases did the target correspond to the target associated with the respective EPIC numbers, and they would thus have been missed 
had only the primary target been extracted.  

\section{Conclusion}
\label{sec:dis}

We have presented our version of a K2 data analysis pipeline, with the objective that it should be fully automatic and work robustly. From the analysis of LC targets from the C0 proposal GO0118 we found the the pipeline indeed works very robustly, and was able to separate close targets and extract data for multiple targets in a given pixel frame. This resulted in an increase in the number of available light curves by a factor of ${\sim}10.4$ for C0, and will naturally vary with the amount of crowding in the different campaigns. Given the large increase in number of potential targets for each assigned EPIC, it needs to be settled how these new targets might be named and identified in other studies.

Concerning the construction of pixel masks we note that many of the published studies of K2 data apply circular masks. But, the flux distribution for a target in K2 is generally far from circular and symmetric, especially if a summed image is used. If a circular mask is used it needs to be large enough to encompass the movement of the target on the CCD; this in turn considerably increases the risk of contamination from other nearby targets. The use of clustering of pixels from the summed image for defining the masks better approximates the actual flux distribution of the target.
For later versions of the pipeline we will investigate in greater detail if any weighting of the pixel masks can lead to a reduction of the high-frequency noise, \eg, as measured via the point-to-point MDV. In relation to this we will also test further the potential impact of a high spatial frequency of the derived pixel masks.    
More effort needs to be invested in improving the correction of instrumental trends via the 2D method. When data from C3 becomes available, where the fine pointing of the spacecraft should be improved, we will revisit this method in more detail. This could also include an implementation of the procedure outlined in \citet[][]{KASCdoc2, KASCdoc1}. We will also continue to try and improve the 1D correction, that in our tests still seems to leave artefacts at harmonics of $\rm {\sim}47.2281\, \mu Hz$. A better removal of these artefacts is clearly needed if an automatic search of asteroseismic power is desired, and simply masking the peaks in the power spectrum will only have a limited impact if the effect of the spectral window is neglected. Our attempt at cleaning the instrumental peaks did improve the power spectrum, but still could not fully remove the instrumental peaks and the window function persisted, which might be expected from cleaning a highly non-sinusoidal signal. As part of the correction we will look into measures other than the centroids for the position of the stars on the CCD; this could include     
the construction of a mean relative movement on the CCD from combining the measures of all targets in a given pixel frame. Also of interest is whether the house-keeping data from the \kp spacecraft can be incorporated for a better overall positional correction.
We will attempt to improve the treatment of saturated targets, which are difficult to deal with via the DBSCAN clustering routine. Aspects that should be improved here are, for instance, a better separation of targets that fall within or close to high-flux pixels from a saturated target.   

We note that our method could potentially be used for dense fields including stellar clusters, and could also be applied to super-stamps from K2 and the nominal \kp mission, as well as the upcoming TESS \citep[][]{2014SPIE.9143E..20R} and PLATO 2.0 \citep[][]{2013arXiv1310.0696R} missions\footnote{TESS: Transiting Exoplanet Survey Satellite; PLATO: PLAnetary Transits and Oscillation of stars.}.
During the development of \pipe we tested the application of the pixel clustering on every time step for the pixel frame of a given target rather than using the summed image.
A complication of this method over using the summed image is that the number of targets identified in the pixel frame varies slightly with time due to noise, and the cluster number of a given targets will also vary in time.
From tests of this version of the pipeline on K2 engineering data, we found that using the pixel clustering on every time step could enable the detection of asteroids and/or comets (or other unidentified objects) as they passed through the pixel frame \citep[see][for an analysis of asteroids found during the K2 engineering run]{2015arXiv150105967S}. When scatter plotting centroid estimates for all identified targets (at a given time step) against time, moving targets such as asteroids make clear centroid trails that deviate from the horizontal trails of quasi-stationary targets such as stars. Identification and analysis of such centroid trails could lead to the detection and tracking of hitherto unknown asteroids/comets.


\section*{Acknowledgments} 
\small{
We would like to thank all active participants at the first K2 data analysis workshop in Aarhus (Denmark, 2014) for many useful discussions on approaches to K2 data extraction and correction. A special thanks to Bram Buysschaert for giving us the idea of cleaning residual peaks from the power spectrum, to Daniel Huber for providing useful input on the EPIC, and to Hans Kjeldsen for commenting on the paper. Finally, we would like to thank the anonymous referee for suggestions and comments that helped to improve the final version of this paper.

Funding for the Stellar Astrophysics Centre (SAC) is provided by The Danish National Research Foundation (Grant agreement no.: DNRF106). The research is supported by the ASTERISK project (ASTERoseismic Investigations with SONG and \kp) funded by the European Research Council (Grant agreement no.: 267864).

W.J.C., G.R.D., and C.D.J. acknowledges the support of the UK Science and Technology Facilities Council (STFC).

This research took advantage of the SIMBAD and VizieR databases at the CDS, Strasbourg (France); NASAs Astrophysics Data System Bibliographic Services (adswww.harvard.edu); arxiv.org, maintained and operated by the Cornell University Library; the USNOFS Image and Catalogue Archive operated by the United States Naval Observatory, Flagstaff Station (\url{http://www.nofs.navy.mil/data/fchpix/}).

}

\vspace{1cm}

\bibliography{MasterBIB.bib}

\begin{thebibliography}{51}
\expandafter\ifx\csname natexlab\endcsname\relax\def\natexlab#1{#1}\fi

\bibitem[{{Aigrain} {et~al.}(2014){Aigrain}, {Hodgkin}, {Irwin}, {Lewis}, \&
  {Roberts}}]{2014arXiv1412.6304A}
{Aigrain}, S., {Hodgkin}, S.~T., {Irwin}, M.~J., {Lewis}, J.~R., \& {Roberts},
  S.~J. 2014, ArXiv e-prints 1412.6304

\bibitem[{{Appourchaux}(2004)}]{2004A&A...428.1039A}
{Appourchaux}, T. 2004, \aap, 428, 1039

\bibitem[{{Ballard} {et~al.}(2010){Ballard}, {Charbonneau}, {Deming},
  {Knutson}, {Christiansen}, {Holman}, {Fabrycky}, {Seager}, \&
  {A'Hearn}}]{2010PASP..122.1341B}
{Ballard}, S., {Charbonneau}, D., {Deming}, D., {et~al.} 2010, \pasp, 122, 1341

\bibitem[{{Basri} {et~al.}(2013){Basri}, {Walkowicz}, \&
  {Reiners}}]{2013ApJ...769...37B}
{Basri}, G., {Walkowicz}, L.~M., \& {Reiners}, A. 2013, \apj, 769, 37

\bibitem[{{Belmonte} {et~al.}(1991){Belmonte}, {Chevreton}, {Mangeney},
  {Praderie}, {Saint-Pe}, {Puget}, {Alvarez}, \& {Roca
  Cortes}}]{1991A&A...246...71B}
{Belmonte}, J.~A., {Chevreton}, M., {Mangeney}, A., {et~al.} 1991, \aap, 246,
  71

\bibitem[{{Beucher} \& {Lantuejoul}(1979)}]{citeulike:2335595}
{Beucher}, S., \& {Lantuejoul}, C. 1979, in International Workshop on Image
  Processing: Real-time Edge and Motion Detection/Estimation, Rennes, France.

\bibitem[{{Beucher} \& {Meyer}(1993)}]{Beucher1993}
{Beucher}, S., \& {Meyer}, F. 1993, Optical Engineering, 34, 433

\bibitem[{{Borucki} {et~al.}(2010){Borucki}, {Koch}, {Basri}, {Batalha},
  {Brown}, {Caldwell}, {Caldwell}, {Christensen-Dalsgaard}, {Cochran},
  {DeVore}, {Dunham}, {Dupree}, {Gautier}, {Geary}, {Gilliland}, {Gould},
  {Howell}, {Jenkins}, {Kondo}, {Latham}, {Marcy}, {Meibom}, {Kjeldsen},
  {Lissauer}, {Monet}, {Morrison}, {Sasselov}, {Tarter}, {Boss}, {Brownlee},
  {Owen}, {Buzasi}, {Charbonneau}, {Doyle}, {Fortney}, {Ford}, {Holman},
  {Seager}, {Steffen}, {Welsh}, {Rowe}, {Anderson}, {Buchhave}, {Ciardi},
  {Walkowicz}, {Sherry}, {Horch}, {Isaacson}, {Everett}, {Fischer}, {Torres},
  {Johnson}, {Endl}, {MacQueen}, {Bryson}, {Dotson}, {Haas}, {Kolodziejczak},
  {Van Cleve}, {Chandrasekaran}, {Twicken}, {Quintana}, {Clarke}, {Allen},
  {Li}, {Wu}, {Tenenbaum}, {Verner}, {Bruhweiler}, {Barnes}, \&
  {Prsa}}]{2010Sci...327..977B}
{Borucki}, W.~J., {Koch}, D., {Basri}, G., {et~al.} 2010, Science, 327, 977

\bibitem[{{Brown} {et~al.}(2014){Brown}, {Anderson}, {Armstrong}, {Bouchy},
  {Collier Cameron}, {Delrez}, {Doyle}, {Gillon}, {Gomez Maqueo Chew}, {Hebb},
  {Hebrard}, {Hellier}, {Jehin}, {Lendl}, {Maxted}, {McCormac},
  {Neveu-VanMalle}, {Pollacco}, {Queloz}, {Segransan}, {Smalley}, {Turner},
  {Triaud}, \& {Udry}}]{2014arXiv1412.7761B}
{Brown}, D.~J.~A., {Anderson}, D.~R., {Armstrong}, D.~J., {et~al.} 2014, ArXiv
  e-prints 1412.7761

\bibitem[{{Brown} {et~al.}(2011){Brown}, {Latham}, {Everett}, \&
  {Esquerdo}}]{2011AJ....142..112B}
{Brown}, T.~M., {Latham}, D.~W., {Everett}, M.~E., \& {Esquerdo}, G.~A. 2011,
  \aj, 142, 112

\bibitem[{{Bryson} {et~al.}(2010){Bryson}, {Tenenbaum}, {Jenkins},
  {Chandrasekaran}, {Klaus}, {Caldwell}, {Gilliland}, {Haas}, {Dotson}, {Koch},
  \& {Borucki}}]{2010ApJ...713L..97B}
{Bryson}, S.~T., {Tenenbaum}, P., {Jenkins}, J.~M., {et~al.} 2010, \apjl, 713,
  L97

\bibitem[{{Calabretta} \& {Greisen}(2002)}]{2002A&A...395.1077C}
{Calabretta}, M.~R., \& {Greisen}, E.~W. 2002, \aap, 395, 1077

\bibitem[{{Christiansen} {et~al.}(2012){Christiansen}, {Jenkins}, {Caldwell},
  {Burke}, {Tenenbaum}, {Seader}, {Thompson}, {Barclay}, {Clarke}, {Li},
  {Smith}, {Stumpe}, {Twicken}, \& {Van Cleve}}]{2012PASP..124.1279C}
{Christiansen}, J.~L., {Jenkins}, J.~M., {Caldwell}, D.~A., {et~al.} 2012,
  \pasp, 124, 1279

\bibitem[{Cleveland(1979)}]{MR556476}
Cleveland, W.~S. 1979, Journal of the American Statistical Association, 74, 829

\bibitem[{Cleveland(1981)}]{lowess}
---. 1981, The American Statistician, 35, 54

\bibitem[{{Davies} {et~al.}(2015){Davies}, {Chaplin}, {Farr}, {Garc{\'{\i}}a},
  {Lund}, {Mathis}, {Metcalfe}, {Appourchaux}, {Basu}, {Benomar}, {Campante},
  {Ceillier}, {Elsworth}, {Handberg}, {Salabert}, \&
  {Stello}}]{2015MNRAS.446.2959D}
{Davies}, G.~R., {Chaplin}, W.~J., {Farr}, W.~M., {et~al.} 2015, \mnras, 446,
  2959

\bibitem[{{Ester} {et~al.}(1996){Ester}, {Kriegel}, {Sander}, \&
  {Xu}}]{ref:dbscan}
{Ester}, M., {Kriegel}, H.-p., {Sander}, J., \& {Xu}, X. 1996, in  (AAAI
  Press), 226--231

\bibitem[{{Foreman-Mackey} {et~al.}(2013){Foreman-Mackey}, {Hogg}, {Lang}, \&
  {Goodman}}]{2013PASP..125..306F}
{Foreman-Mackey}, D., {Hogg}, D.~W., {Lang}, D., \& {Goodman}, J. 2013, \pasp,
  125, 306

\bibitem[{{Fraquelli} \& {Thompson}(2012)}]{kepman}
{Fraquelli}, D., \& {Thompson}, S.~E. 2012, Kepler Archive Manual,
  (KDMC-10008-004)

\bibitem[{{Gilliland} {et~al.}(2010{\natexlab{a}}){Gilliland}, {Jenkins},
  {Borucki}, {Bryson}, {Caldwell}, {Clarke}, {Dotson}, {Haas}, {Hall}, {Klaus},
  {Koch}, {McCauliff}, {Quintana}, {Twicken}, \& {van
  Cleve}}]{2010ApJ...713L.160G}
{Gilliland}, R.~L., {Jenkins}, J.~M., {Borucki}, W.~J., {et~al.}
  2010{\natexlab{a}}, \apjl, 713, L160

\bibitem[{{Gilliland} {et~al.}(2010{\natexlab{b}}){Gilliland}, {Brown},
  {Christensen-Dalsgaard}, {Kjeldsen}, {Aerts}, {Appourchaux}, {Basu},
  {Bedding}, {Chaplin}, {Cunha}, {De Cat}, {De Ridder}, {Guzik}, {Handler},
  {Kawaler}, {Kiss}, {Kolenberg}, {Kurtz}, {Metcalfe}, {Monteiro}, {Szab{\'o}},
  {Arentoft}, {Balona}, {Debosscher}, {Elsworth}, {Quirion}, {Stello},
  {Su{\'a}rez}, {Borucki}, {Jenkins}, {Koch}, {Kondo}, {Latham}, {Rowe}, \&
  {Steffen}}]{2010PASP..122..131G}
{Gilliland}, R.~L., {Brown}, T.~M., {Christensen-Dalsgaard}, J., {et~al.}
  2010{\natexlab{b}}, \pasp, 122, 131

\bibitem[{{Gilliland} {et~al.}(2011){Gilliland}, {Chaplin}, {Dunham},
  {Argabright}, {Borucki}, {Basri}, {Bryson}, {Buzasi}, {Caldwell}, {Elsworth},
  {Jenkins}, {Koch}, {Kolodziejczak}, {Miglio}, {van Cleve}, {Walkowicz}, \&
  {Welsh}}]{2011ApJS..197....6G}
{Gilliland}, R.~L., {Chaplin}, W.~J., {Dunham}, E.~W., {et~al.} 2011, \apjs,
  197, 6

\bibitem[{Ginsburg {et~al.}(2013)Ginsburg, Robitaille, Parikh, Deil, Mirocha,
  Woillez, Svoboda, Willett, Allen, Grollier, Persson, \& Shiga}]{Ginsburg2013}
Ginsburg, A., Robitaille, T., Parikh, M., {et~al.} 2013

\bibitem[{{Greisen} \& {Calabretta}(2002)}]{2002A&A...395.1061G}
{Greisen}, E.~W., \& {Calabretta}, M.~R. 2002, \aap, 395, 1061

\bibitem[{{Greisen} {et~al.}(2006){Greisen}, {Calabretta}, {Valdes}, \&
  {Allen}}]{2006A&A...446..747G}
{Greisen}, E.~W., {Calabretta}, M.~R., {Valdes}, F.~G., \& {Allen}, S.~L. 2006,
  \aap, 446, 747

\bibitem[{{Handberg} \& {Lund}(2014)}]{2014MNRAS.445.2698H}
{Handberg}, R., \& {Lund}, M.~N. 2014, \mnras, 445, 2698

\bibitem[{{Howell} {et~al.}(2012){Howell}, {Rowe}, {Bryson}, {Quinn}, {Marcy},
  {Isaacson}, {Ciardi}, {Chaplin}, {Metcalfe}, {Monteiro}, {Appourchaux},
  {Basu}, {Creevey}, {Gilliland}, {Quirion}, {Stello}, {Kjeldsen},
  {Christensen-Dalsgaard}, {Elsworth}, {Garc{\'{\i}}a}, {Houdek}, {Karoff},
  {Molenda-{\.Z}akowicz}, {Thompson}, {Verner}, {Torres}, {Fressin}, {Crepp},
  {Adams}, {Dupree}, {Sasselov}, {Dressing}, {Borucki}, {Koch}, {Lissauer},
  {Latham}, {Buchhave}, {Gautier}, {Everett}, {Horch}, {Batalha}, {Dunham},
  {Szkody}, {Silva}, {Mighell}, {Holberg}, {Ballot}, {Bedding}, {Bruntt},
  {Campante}, {Handberg}, {Hekker}, {Huber}, {Mathur}, {Mosser}, {R{\'e}gulo},
  {White}, {Christiansen}, {Middour}, {Haas}, {Hall}, {Jenkins}, {McCaulif},
  {Fanelli}, {Kulesa}, {McCarthy}, \& {Henze}}]{2012ApJ...746..123H}
{Howell}, S.~B., {Rowe}, J.~F., {Bryson}, S.~T., {et~al.} 2012, \apj, 746, 123

\bibitem[{{Howell} {et~al.}(2014){Howell}, {Sobeck}, {Haas}, {Still},
  {Barclay}, {Mullally}, {Troeltzsch}, {Aigrain}, {Bryson}, {Caldwell},
  {Chaplin}, {Cochran}, {Huber}, {Marcy}, {Miglio}, {Najita}, {Smith},
  {Twicken}, \& {Fortney}}]{2014PASP..126..398H}
{Howell}, S.~B., {Sobeck}, C., {Haas}, M., {et~al.} 2014, \pasp, 126, 398

\bibitem[{{Jenkins} {et~al.}(2010){Jenkins}, {Caldwell}, {Chandrasekaran},
  {Twicken}, {Bryson}, {Quintana}, {Clarke}, {Li}, {Allen}, {Tenenbaum}, {Wu},
  {Klaus}, {Middour}, {Cote}, {McCauliff}, {Girouard}, {Gunter}, {Wohler},
  {Sommers}, {Hall}, {Uddin}, {Wu}, {Bhavsar}, {Van Cleve}, {Pletcher},
  {Dotson}, {Haas}, {Gilliland}, {Koch}, \& {Borucki}}]{2010ApJ...713L..87J}
{Jenkins}, J.~M., {Caldwell}, D.~A., {Chandrasekaran}, H., {et~al.} 2010,
  \apjl, 713, L87

\bibitem[{{Kjeldsen} {et~al.}(2013{\natexlab{a}}){Kjeldsen}, {Arentoft}, \&
  {Christensen-Dalsgaard}}]{KASCdoc2}
{Kjeldsen}, H., {Arentoft}, T., \& {Christensen-Dalsgaard}, J.
  2013{\natexlab{a}}, KASC document, DASC/KASOC/0043,
  \url{http://astro.phys.au.dk/~hans/Call\_for\_White\_Paper/DASC_KASOC_0043_2%
.pdf}

\bibitem[{{Kjeldsen} {et~al.}(2013{\natexlab{b}}){Kjeldsen}, {Arentoft}, \&
  {Christensen-Dalsgaard}}]{KASCdoc1}
---. 2013{\natexlab{b}}, KASC document, DASC/KASOC/0044,
  \url{http://astro.phys.au.dk/~hans/Call_for_White_Paper/DASC_KASOC_0044_1.pd%
f}

\bibitem[{{Knutson} {et~al.}(2008){Knutson}, {Charbonneau}, {Allen}, {Burrows},
  \& {Megeath}}]{2008ApJ...673..526K}
{Knutson}, H.~A., {Charbonneau}, D., {Allen}, L.~E., {Burrows}, A., \&
  {Megeath}, S.~T. 2008, \apj, 673, 526

\bibitem[{{Koch} {et~al.}(2010){Koch}, {Borucki}, {Basri}, {Batalha}, {Brown},
  {Caldwell}, {Christensen-Dalsgaard}, {Cochran}, {DeVore}, {Dunham},
  {Gautier}, {Geary}, {Gilliland}, {Gould}, {Jenkins}, {Kondo}, {Latham},
  {Lissauer}, {Marcy}, {Monet}, {Sasselov}, {Boss}, {Brownlee}, {Caldwell},
  {Dupree}, {Howell}, {Kjeldsen}, {Meibom}, {Morrison}, {Owen}, {Reitsema},
  {Tarter}, {Bryson}, {Dotson}, {Gazis}, {Haas}, {Kolodziejczak}, {Rowe}, {Van
  Cleve}, {Allen}, {Chandrasekaran}, {Clarke}, {Li}, {Quintana}, {Tenenbaum},
  {Twicken}, \& {Wu}}]{2010ApJ...713L..79K}
{Koch}, D.~G., {Borucki}, W.~J., {Basri}, G., {et~al.} 2010, \apjl, 713, L79

\bibitem[{{Lucas}(1977)}]{GLucas1977}
{Lucas}, G.~W. 1977, A new hope,
  \url{http://en.wikipedia.org/wiki/1138_%28number%29}

\bibitem[{{Lund} {et~al.}(2012){Lund}, {Chaplin}, \&
  {Kjeldsen}}]{2012MNRAS.427.1784L}
{Lund}, M.~N., {Chaplin}, W.~J., \& {Kjeldsen}, H. 2012, \mnras, 427, 1784

\bibitem[{{Lund} {et~al.}(2014){Lund}, {Kjeldsen}, {Christensen-Dalsgaard},
  {Handberg}, \& {Silva Aguirre}}]{2014ApJ...782....2L}
{Lund}, M.~N., {Kjeldsen}, H., {Christensen-Dalsgaard}, J., {Handberg}, R., \&
  {Silva Aguirre}, V. 2014, \apj, 782, 2

\bibitem[{{Monet} {et~al.}(2003){Monet}, {Levine}, {Canzian}, {Ables}, {Bird},
  {Dahn}, {Guetter}, {Harris}, {Henden}, {Leggett}, {Levison}, {Luginbuhl},
  {Martini}, {Monet}, {Munn}, {Pier}, {Rhodes}, {Riepe}, {Sell}, {Stone},
  {Vrba}, {Walker}, {Westerhout}, {Brucato}, {Reid}, {Schoening}, {Hartley},
  {Read}, \& {Tritton}}]{2003AJ....125..984M}
{Monet}, D.~G., {Levine}, S.~E., {Canzian}, B., {et~al.} 2003, \aj, 125, 984

\bibitem[{{Pedregosa} {et~al.}(2011){Pedregosa}, {Varoquaux}, {Gramfort},
  {Michel}, {Thirion}, {Grisel}, {Blondel}, {Prettenhofer}, {Weiss}, {Dubourg},
  {Vanderplas}, {Passos}, {Cournapeau}, {Brucher}, {Perrot}, \&
  {Duchesnay}}]{paper:scikit-learn}
{Pedregosa}, F., {Varoquaux}, G., {Gramfort}, A., {et~al.} 2011, Journal of
  Machine Learning Research, 12, 2825

\bibitem[{{Pinsonneault} {et~al.}(2014){Pinsonneault}, {Elsworth}, {Epstein},
  {Hekker}, {M{\'e}sz{\'a}ros}, {Chaplin}, {Johnson}, {Garc{\'{\i}}a},
  {Holtzman}, {Mathur}, {Garc{\'{\i}}a P{\'e}rez}, {Silva Aguirre}, {Girardi},
  {Basu}, {Shetrone}, {Stello}, {Allende Prieto}, {An}, {Beck}, {Beers},
  {Bizyaev}, {Bloemen}, {Bovy}, {Cunha}, {De Ridder}, {Frinchaboy},
  {Garc{\'{\i}}a-Hern{\'a}ndez}, {Gilliland}, {Harding}, {Hearty}, {Huber},
  {Ivans}, {Kallinger}, {Majewski}, {Metcalfe}, {Miglio}, {Mosser}, {Muna},
  {Nidever}, {Schneider}, {Serenelli}, {Smith}, {Tayar}, {Zamora}, \&
  {Zasowski}}]{2014ApJS..215...19P}
{Pinsonneault}, M.~H., {Elsworth}, Y., {Epstein}, C., {et~al.} 2014, \apjs,
  215, 19

\bibitem[{{Ponman}(1981)}]{1981MNRAS.196..583P}
{Ponman}, T. 1981, \mnras, 196, 583

\bibitem[{{Rauer} {et~al.}(2013){Rauer}, {Catala}, {Aerts}, {Appourchaux},
  {Benz}, {Brandeker}, {Christensen-Dalsgaard}, {Deleuil}, {Gizon}, {Goupil},
  {G{\"u}del}, {Janot-Pacheco}, {Mas-Hesse}, {Pagano}, {Piotto}, {Pollacco},
  {Santos}, {Smith}, {Su{\'a}rez}, {Szab{\'o}}, {Udry}, {Adibekyan}, {Alibert},
  {Almenara}, {Amaro-Seoane}, {Ammler-von Eiff}, {Asplund}, {Antonello},
  {Ball}, {Barnes}, {Baudin}, {Belkacem}, {Bergemann}, {Bihain}, {Birch},
  {Bonfils}, {Boisse}, {Bonomo}, {Borsa}, {Brand{\~a}o}, {Brocato}, {Brun},
  {Burleigh}, {Burston}, {Cabrera}, {Cassisi}, {Chaplin}, {Charpinet},
  {Chiappini}, {Church}, {Csizmadia}, {Cunha}, {Damasso}, {Davies}, {Deeg},
  {D{\'I}az}, {Dreizler}, {Dreyer}, {Eggenberger}, {Ehrenreich},
  {Eigm{\"u}ller}, {Erikson}, {Farmer}, {Feltzing}, {de Oliveira Fialho},
  {Figueira}, {Forveille}, {Fridlund}, {Garc{\'i}a}, {Giommi}, {Giuffrida},
  {Godolt}, {Gomes da Silva}, {Granzer}, {Grenfell}, {Grotsch-Noels},
  {G{\"u}nther}, {Haswell}, {Hatzes}, {H{\'e}brard}, {Hekker}, {Helled},
  {Heng}, {Jenkins}, {Johansen}, {Khodachenko}, {Kislyakova}, {Kley}, {Kolb},
  {Krivova}, {Kupka}, {Lammer}, {Lanza}, {Lebreton}, {Magrin}, {Marcos-Arenal},
  {Marrese}, {Marques}, {Martins}, {Mathis}, {Mathur}, {Messina}, {Miglio},
  {Montalban}, {Montalto}, {Monteiro}, {Moradi}, {Moravveji}, {Mordasini},
  {Morel}, {Mortier}, {Nascimbeni}, {Nelson}, {Nielsen}, {Noack}, {Norton},
  {Ofir}, {Oshagh}, {Ouazzani}, {P{\'a}pics}, {Parro}, {Petit}, {Plez},
  {Poretti}, {Quirrenbach}, {Ragazzoni}, {Raimondo}, {Rainer}, {Reese},
  {Redmer}, {Reffert}, {Rojas-Ayala}, {Roxburgh}, {Salmon}, {Santerne},
  {Schneider}, {Schou}, {Schuh}, {Schunker}, {Silva-Valio}, {Silvotti},
  {Skillen}, {Snellen}, {Sohl}, {Sousa}, {Sozzetti}, {Stello}, {Strassmeier},
  {\v{S}vanda}, {Szab{\'o}}, {Tkachenko}, {Valencia}, {van Grootel},
  {Vauclair}, {Ventura}, {Wagner}, {Walton}, {Weingrill}, {Werner}, {Wheatley},
  \& {Zwintz}}]{2013arXiv1310.0696R}
{Rauer}, H., {Catala}, C., {Aerts}, C., {et~al.} 2013, ArXiv e-prints 1310.0696

\bibitem[{{Ricker} {et~al.}(2014){Ricker}, {Winn}, {Vanderspek}, {Latham},
  {Bakos}, {Bean}, {Berta-Thompson}, {Brown}, {Buchhave}, {Butler}, {Butler},
  {Chaplin}, {Charbonneau}, {Christensen-Dalsgaard}, {Clampin}, {Deming},
  {Doty}, {De Lee}, {Dressing}, {Dunham}, {Endl}, {Fressin}, {Ge}, {Henning},
  {Holman}, {Howard}, {Ida}, {Jenkins}, {Jernigan}, {Johnson}, {Kaltenegger},
  {Kawai}, {Kjeldsen}, {Laughlin}, {Levine}, {Lin}, {Lissauer}, {MacQueen},
  {Marcy}, {McCullough}, {Morton}, {Narita}, {Paegert}, {Palle}, {Pepe},
  {Pepper}, {Quirrenbach}, {Rinehart}, {Sasselov}, {Sato}, {Seager},
  {Sozzetti}, {Stassun}, {Sullivan}, {Szentgyorgyi}, {Torres}, {Udry}, \&
  {Villasenor}}]{2014SPIE.9143E..20R}
{Ricker}, G.~R., {Winn}, J.~N., {Vanderspek}, R., {et~al.} 2014, in Society of
  Photo-Optical Instrumentation Engineers (SPIE) Conference Series, Vol. 9143,
  Society of Photo-Optical Instrumentation Engineers (SPIE) Conference Series,
  20

\bibitem[{{Scargle}(1982)}]{1982ApJ...263..835S}
{Scargle}, J.~D. 1982, \apj, 263, 835

\bibitem[{Scott(1979)}]{SCOTT01121979}
Scott, D.~W. 1979, Biometrika, 66, 605

\bibitem[{{Skrutskie} {et~al.}(2006){Skrutskie}, {Cutri}, {Stiening},
  {Weinberg}, {Schneider}, {Carpenter}, {Beichman}, {Capps}, {Chester},
  {Elias}, {Huchra}, {Liebert}, {Lonsdale}, {Monet}, {Price}, {Seitzer},
  {Jarrett}, {Kirkpatrick}, {Gizis}, {Howard}, {Evans}, {Fowler}, {Fullmer},
  {Hurt}, {Light}, {Kopan}, {Marsh}, {McCallon}, {Tam}, {Van Dyk}, \&
  {Wheelock}}]{2006AJ....131.1163S}
{Skrutskie}, M.~F., {Cutri}, R.~M., {Stiening}, R., {et~al.} 2006, \aj, 131,
  1163

\bibitem[{Spratling \& Mortari(2009)}]{a2010093}
Spratling, B.~B., \& Mortari, D. 2009, Algorithms, 2, 93

\bibitem[{{Stevenson} {et~al.}(2012){Stevenson}, {Harrington}, {Fortney},
  {Loredo}, {Hardy}, {Nymeyer}, {Bowman}, {Cubillos}, {Bowman}, \&
  {Hardin}}]{2012ApJ...754..136S}
{Stevenson}, K.~B., {Harrington}, J., {Fortney}, J.~J., {et~al.} 2012, \apj,
  754, 136

\bibitem[{{Szab{\'o}} {et~al.}(2015){Szab{\'o}}, {S{\'a}rneczky}, {Szab{\'o}},
  {P{\'a}l}, {Kiss}, {Cs{\'a}k}, {Ill{\'e}s}, {R{\'a}cz}, \&
  {Kiss}}]{2015arXiv150105967S}
{Szab{\'o}}, R., {S{\'a}rneczky}, K., {Szab{\'o}}, G.~M., {et~al.} 2015, ArXiv
  e-prints 1501.05967

\bibitem[{{Van Der Walt} {et~al.}(2014){Van Der Walt}, {Sch\"{o}nberger},
  {Nunez-Iglesias}, {Boulogne}, {Warner}, {Yager}, {Gouillart}, \&
  {Yu}}]{ref:scikit-image}
{Van Der Walt}, S., {Sch\"{o}nberger}, J.~L., {Nunez-Iglesias}, J., {et~al.}
  2014, PeerJ, 2014

\bibitem[{{Vanderburg}(2014)}]{2014arXiv1412.1827V}
{Vanderburg}, A. 2014, ArXiv e-prints 1412.1827

\bibitem[{{Vanderburg} \& {Johnson}(2014)}]{2014PASP..126..948V}
{Vanderburg}, A., \& {Johnson}, J.~A. 2014, \pasp, 126, 948

\end{thebibliography}

\end{document}